\documentclass[
reprint,
showkeys,
longbibliography,
nofootinbib,
superscriptaddress,
aps,
notitlepage
]{revtex4-1}
\usepackage[utf8]{inputenc}

\usepackage{amsmath}
\usepackage{amssymb}
\usepackage{bm}
\usepackage{enumitem}
\usepackage[acronym]{glossaries}
\usepackage{graphicx}
\usepackage[colorlinks,unicode]{hyperref}
\usepackage[capitalise]{cleveref}  
\usepackage{mathtools}
\usepackage{multirow}
\usepackage{physics}
\usepackage{siunitx}
\usepackage[dvipsnames]{xcolor}

\DeclareSIUnit\week{week}

\makeglossaries
\newacronym{dm}{DM}{dark matter}
\newacronym{ce}{CE}{Cosmic Explorer}
\newacronym{et}{ET}{Einstein Telescope}
\newacronym{snr}{SNR}{signal-to-noise ratio}

\makeatletter
\renewcommand\onecolumngrid{%
\do@columngrid{one}{\@ne}%
\def\set@footnotewidth{\onecolumngrid}%
\def\footnoterule{\kern-6pt\hrule width 1.5in\kern6pt}%
}
\makeatother

\newcommand\myshade{80}
\colorlet{mylinkcolor}{ForestGreen}
\colorlet{mycitecolor}{Red}
\colorlet{myurlcolor}{violet}

\hypersetup{
  linkcolor  = mylinkcolor!\myshade!black,
  citecolor  = mycitecolor!\myshade!black,
  urlcolor   = myurlcolor!\myshade!black,
  colorlinks = true
}

\newcommand{\rmDM}{\mathrm{DM}}

\newcommand{\diff}{\mathrm{d}}
\newcommand{\fPBH}{f_\mathrm{PBH}}

\DeclareSIUnit\solarmass{\ensuremath{\mathrm{M}_\odot}}
\DeclareSIUnit\parsec{pc}
\DeclareSIUnit\year{yr}

\newcommand{\GRAPPA}{Gravitation Astroparticle Physics Amsterdam (GRAPPA),\\ Institute for Theoretical Physics Amsterdam and Delta Institute for Theoretical Physics,\\ University of Amsterdam, Science Park 904, 1098 XH Amsterdam, The Netherlands}
\newcommand{\UdeM}{Département de Physique, Université de Montréal, 1375 Avenue Thérèse-Lavoie-Roux, Montréal, QC H2V 0B3, Canada}
\newcommand{\Mila}{Mila -- Quebec AI Institute, 6666 St-Urbain, \#200, Montreal, QC, H2S 3H1}
\newcommand{\IFCA}{Instituto de F\'isica de Cantabria (IFCA, UC-CSIC), Av.~de Los Castros s/n, 39005 Santander, Spain}

\begin{document}

\title{Measuring dark matter spikes around primordial black holes with Einstein Telescope and Cosmic Explorer}

\author{Philippa S. Cole}
\email{p.s.cole@uva.nl}
\affiliation{\GRAPPA}

\author{Adam Coogan}
\email{adam.coogan@umontreal.ca}
\affiliation{\GRAPPA}
\affiliation{\UdeM}
\affiliation{\Mila}

\author{Bradley J. Kavanagh}
\email{kavanagh@ifca.unican.es}
\affiliation{\IFCA}

\author{Gianfranco Bertone}
\email{g.bertone@uva.nl}
\affiliation{\GRAPPA}

\begin{abstract}
    Future ground-based gravitational wave observatories will be ideal probes of the environments surrounding black holes with masses $\SIrange{1}{10}{\solarmass}$. Binary black hole mergers with mass ratios of order $q=m_2/m_1\lesssim10^{-3}$ can remain in the frequency band of such detectors for months or years, enabling precision searches for modifications of their gravitational waveforms with respect to vacuum inspirals. As a concrete example of an environmental effect, we consider here a population of binary primordial black holes which are expected to be embedded in dense cold dark matter spikes. We provide a viable formation scenario for these systems compatible with all observational constraints, and predict upper and lower limits on the merger rates of small mass ratio pairs. Given a detected signal of one such system by either Einstein Telescope or Cosmic Explorer, we show that the properties of the binary and of the dark matter spike can be measured to excellent precision with one week's worth of data, if the effect of the dark matter spike on the waveform is taken into account. However, we show that there is a risk of biased parameter inference or missing the events entirely if the effect of the predicted dark matter overdensity around these objects is not properly accounted for.
\end{abstract}


\maketitle

\section{Introduction}

Future ground-based gravitational wave observatories will be capable of detecting binary black hole mergers at unprecedented distances~\cite{Hall_LIGOdoc}. Whilst event rates will be dominated by close-to-equal mass mergers, the enhanced sensitivity and wider frequency range of the proposed Einstein Telescope (ET)~\cite{Punturo:2010zz,Maggiore:2019uih} and Cosmic Explorer (CE)~\cite{Evans:2021gyd} observatories will also open the door to observing intermediate mass ratio mergers. 

The planned frequency ranges of Cosmic Explorer, $f\sim[5,5000]\,{\rm Hz}$, and Einstein Telescope $f\sim[1,5000]\,{\rm Hz}$ mean that the observatories will be sensitive to long duration signals from light systems with mass ratios of order $q=m_2/m_1\sim10^{-3}$, where $m_1$ is the mass of the central compact object and $m_2$ is the mass of its lighter companion. Binaries with $m_1=1-10\,{\rm M_\odot}$ will remain in band for many months or years. Detecting such systems would have exciting consequences for gravitational wave astronomy as well as fundamental physics \cite{Barack:2018yly,Bertone:2019irm}.

Systems with small mass-ratios are particular interesting because they are more likely to be influenced by environmental effects. The gravitational waveform of a binary inspiralling through an environment will be different to that of the equivalent system merging in vacuum. In the case of an environment of collisionless matter, dynamical friction, accretion, and a varying mass enclosed within the orbit alter the dynamics of the binary~\cite{Eda1,Eda2,Macedo:2013qea,Barausse:2014tra,Barausse:2014pra,Yue:2017iwc,Cardoso:2019rou,Hannuksela:2019vip,Kavanagh:2020cfn,Coogan:2021uqv}.
This appears as a gradual change in the cumulative phase of the waveform with respect to the system in vacuum, detectable if a very large number of cycles are observed. 
Dense environments are more likely to survive around small mass-ratio systems, unlike in equal-mass binaries, where any environments are likely to be disrupted on a timescale of a few orbits~\cite{Merritt:2002vj,Kavanagh:2018ggo}.
Given the amount of time these systems will spend in band, this places Einstein Telescope and Cosmic Explorer well to detect this \textit{dephasing} effect in small mass-ratio binaries.

In this work, we explore how well ET and CE can measure the properties of primordial black hole (PBH) binaries embedded in cold dark matter (DM) spikes. PBHs may be formed shortly after the end of inflation from large density fluctuations, contributing to the non-baryonic content of the Universe~\cite{Green:2020jor}.
However, PBHs cannot make up all of the dark matter by themselves in the stellar mass range (see e.g.~\cite{Clessecluster,Youngcluster} for caveats), and therefore must be accompanied by another dark component. If the remaining DM is made up of cold, collisionless particles, it will form dense spikes around primordial black holes with a well-defined density profile~\cite{Bertschinger:1985pd,Mack:2006gz,Ricotti:2007jk,Boudaud:2021irr}, which will have an effect on a PBH binary's dynamics. Intermediate or extreme mass ratio PBH binaries can therefore only be found and correctly interpreted if the effect of the dark matter spike is taken into account. Moreover, electromagnetic signatures of DM spikes around PBHs are largely ruled out for $\fPBH \gtrsim 10^{-8}$~\cite{Lacki:2010zf,Adamek:2019gns,Bertone:2019vsk,Carr:2020mqm}, making gravitational wave searches a particularly promising avenue for discovery of such mixed DM scenarios. 




We show that Einstein Telescope and Cosmic Explorer are ideally positioned to observe gravitational wave signals from dark matter-dressed PBH binaries. In particular, the range of GW frequencies accessible to both experiments makes them sensitive to solar and sub-solar mass PBH binaries in the local Universe~(\cref{sec:reach}). We provide a concrete formation scenario for such PBH binaries and predict upper and lower limits on the merger rates of intermediate-mass-ratio pairs~ (\cref{sec:mergerrate}). We specify the properties of the DM spikes which are expected to form around PBHs and describe how the influence of these DM spikes on the gravitational waveform can be modelled~(\cref{sec:dephasing}). With these tools, we show that it will be possible to distinguish these systems from GR-in-vacuum inspirals and to measure the properties of the binary and the dark matter spike~(\cref{sec:measurability,sec:results}). This is only possible if parameter estimation is conducted with waveform templates that take into account the effects of dynamical friction; we risk missing these signals in the data if it is assumed that all binaries are inspiralling in vacuum. We conclude by discussing these challenges associated with realistic search and inference strategies, as well as relevance of our results to other environmental effects around BH binaires~(\cref{sec:conclusions}).

\section{Einstein Telescope and Cosmic Explorer reach}
\label{sec:reach}

Firstly, we assess which systems are best-placed to be detected by various observatories.

\begin{figure*}
    \centering
    \includegraphics{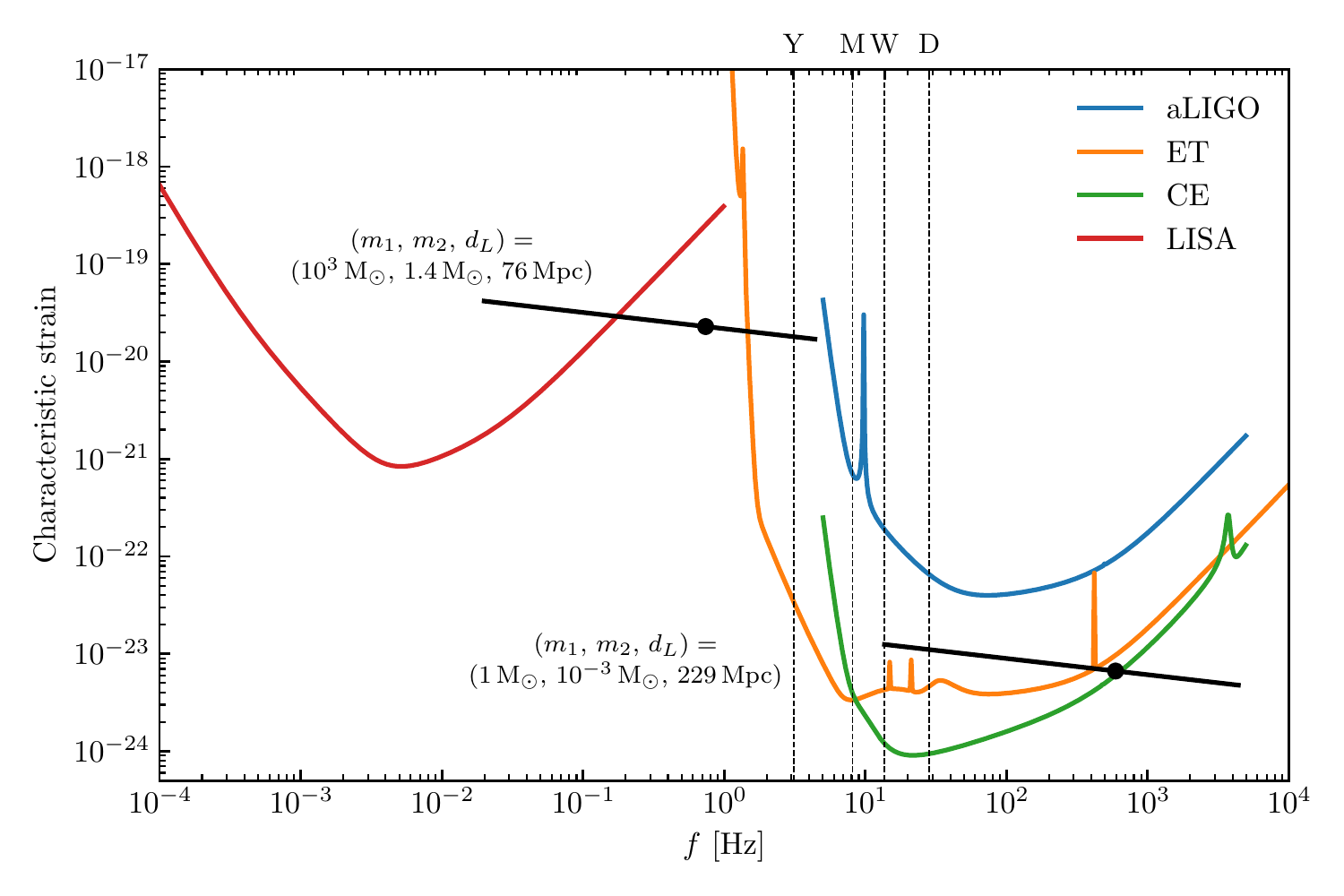}
    \caption{\textbf{Characteristic strain as a function of frequency for the noise in various detectors as well as two primordial dark dress systems.} The heavy and light systems ``trajectories'' (thick black lines) begin five years and one week before the innermost stable circular orbit frequency. The systems' distances were respectively chosen to given signal-to-noise ratios of 15 at LISA and 12 at Cosmic Explorer over these observing durations. The vertical dashed lines indicate the frequencies at a year, month, week and day before coalescence for the light binary. The black dots indicate the frequency above which depletion of the spike via dynamical friction from the compact object becomes negligible (see \cref{sec:dephasing}). The system in the LISA band was studied in detail in Ref.~\cite{Coogan:2021uqv}.}
    \label{fig:trajectories-and-noise}
\end{figure*}

The improvement in sensitivity and frequency reach of Einstein Telescope and Cosmic Explorer beyond aLIGO is shown by the noise power spectral density curves $S_n(t)$ in Fig.~\ref{fig:trajectories-and-noise}. Throughout this paper, we choose our `benchmark' PBH binary to have masses $m_1=1\,{\rm M_\odot}$ and $m_2=10^{-3}\,{\rm M_\odot}$. The characteristic strain~\cite{Moore:2014lga} of the gravitational wave signal caused by the inspiralling of black holes with these masses is shown by the rightmost black line in \cref{fig:trajectories-and-noise}, for one week's worth of frequency evolution. This system lies very nicely in the frequency range of the future ground-based detectors. The presence of a dense DM spike around the heavier PBH will influence the precise frequency evolution of the system, while inspiral of the binary will lead to a gradual depletion of the spike due to feedback effects~\cite{Kavanagh:2020cfn}.

The black dot on each trajectory in \cref{fig:trajectories-and-noise} marks the `break frequency' $f_b$; at frequencies below $f_b$ the timescale for the depletion of the DM spike is much shorter than the timescale for inspiral due to GW emission, while at frequencies above $f_b$ the evolution of the DM spike becomes negligible. This parametrization in terms of $f_b$ is relevant for calculating the effects of feedback on the spike and will be discussed later in  \cref{sec:dephasing}.  As argued in Ref.~\cite{Coogan:2021uqv}, the important point is that in order to detect the dephasing of the gravitational waveform, the break frequency should lie in a region of high sensitivity of the detector. In addition, the system should remain in band for a long time period so that as many dephased cycles as possible are observed. For example, for this benchmark system with a year of observations, approximately $10^7$ cycles occur in band.

For comparison, we also show the LISA noise curve which lies at far lower frequencies, making it more suited to observing the higher-mass systems which have previously been studied in the context of DM-induced dephasing~\cite{Eda1,Eda2,Yue:2017iwc,Yue:2018vtk,Hannuksela:2019vip,Cardoso:2019rou,Kavanagh:2020cfn,Dai:2021olt,Li:2021pxf,Coogan:2021uqv,Becker:2021ivq,Speeney:2022ryg}. We also plot the characteristic strain for five years worth of frequency evolution of the system with $m_1=10^3\,{\rm M_\odot}$, $m_2=1.4\,{\rm M_\odot}$ embedded in a dark matter spike, which was explored in~\cite{Coogan:2021uqv}. It was shown that the properties of the dark matter spike could be reconstructed to very good accuracy with five years of observations, and motivates this investigation into what future ground-based observatories can do for dark matter spike searches.

For each detector, the distance at which our benchmark system will be detectable varies. We calculate the SNR averaged over sky position, orientation and polarization angle as a function of chirp mass $\mathcal{M}=(m_1m_2)^\frac{3}{5}/(m_1+m_2)^\frac{1}{5}$ and luminosity distance to source $d_L$ via
\begin{equation}
\label{eq:SNR}
\mathrm{SNR} = \frac{1}{d_L}\sqrt{4 \int \frac{4}{5}\frac{ h_0^2(f)}{ S_n(f)} \mathrm{d}f} \, ,
\end{equation}
where $S_n(f)$ is the power spectral density for a detector and the amplitude of the Fourier transform of the gravitational waveform $h_0$ is constructed from the following expressions for the phase $\Phi$ and its derivatives with respect to time~\cite{Maggiore:2007ulw}:
\begin{align}
    \begin{split}
    h_0(f) &= \frac{1}{2} \frac{4 \pi^{2/3} G^{5/3} \mathcal{M}^{5/3} f^{2/3}}{c^4} \sqrt{\frac{2\pi}{\ddot{\Phi}}}\,,\\
    \ddot{\Phi} &= 4\pi^2 f \qty(\dv{\Phi}{f})^{-1}\,,\\
    \Phi(f) &= \int_f^{f_\mathrm{ISCO}} \frac{\mathrm{d}t}{\mathrm{d}f^\prime} f^\prime \,\mathrm{d}f^\prime\,,\\
    t(f) &= \frac{1}{2\pi} \int^f \frac{\dd{f'}}{f'} \dv{\Phi}{f}\,.
    \label{eq:strain_and_phase}
    \end{split}
\end{align}
Here $G$ is Newton's gravitational constant, $c$ is the speed of light, and $f_\mathrm{ISCO}$ is the GW frequency at the innermost stable circular orbit (ISCO), which we take to be the point of merger.

\begin{figure*}
    \centering
    \includegraphics[width=\textwidth]{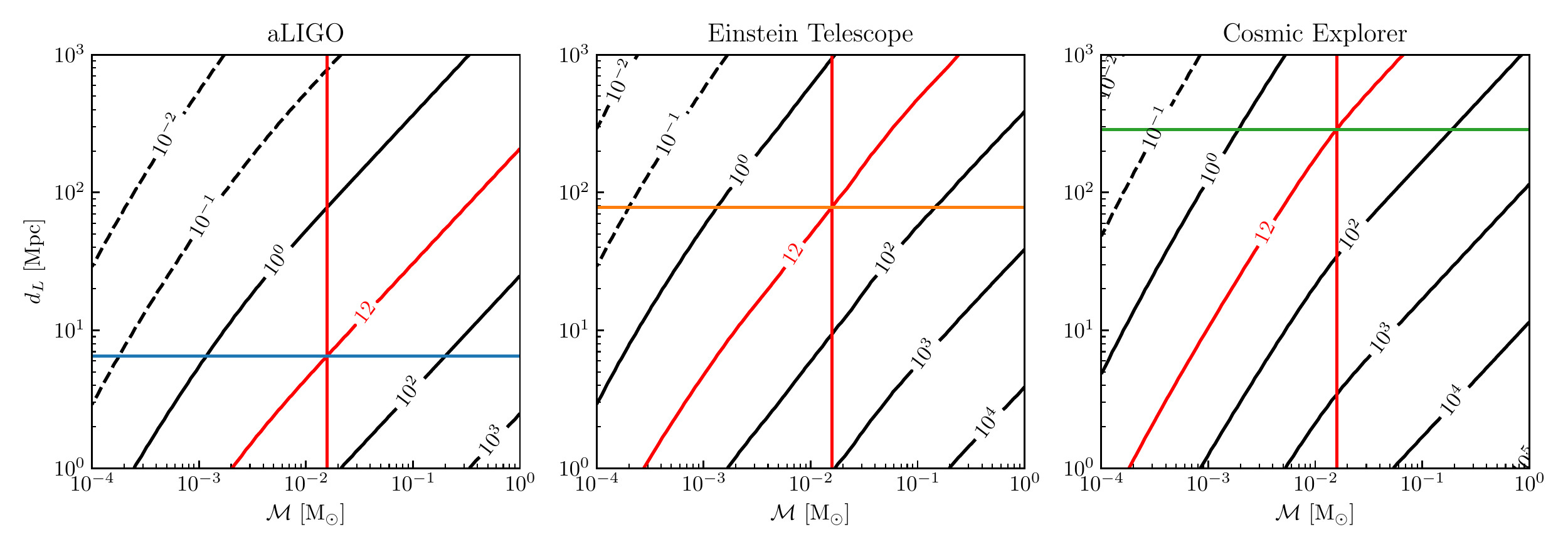}
    \caption{\textbf{\Gls*{snr}s for benchmark system observed by terrestrial detectors.} The vertical red line indicates the chirp mass of our benchmark $(m_1, m_2) = (1, 10^{-3})\si{\solarmass}$ system. The horizontal lines indicate the distances at which this system would have an (optimal) \gls*{snr} of 12 in each detector, with their colors matching the ones used in \Cref{fig:trajectories-and-noise}. The diagonal contours indicate curves of constant \gls*{snr}, with the red one highlighting the special value of 12 which we use to define a system as detectable. Note that the \gls*{snr} for vacuum binaries and the dark dresses we consider is the same since it is not substantially altered by the \gls*{dm} spike. Since a vacuum binary's frequency at merger is not purely a function of its chirp mass, in this plot we assume each system is observed for \SI{1}{\week} up to the high-frequency cutoff for each detector.}
    \label{fig:snrs}
\end{figure*}

The SNRs for aLIGO, ET and CE are shown in \cref{fig:snrs} where we highlight an SNR threshold of 12 as our definition of detectability, although note that this threshold could be lowered if multiple detectors are online simultaneously. With aLIGO, our benchmark system would only be detectable within a small volume of $<300\,{\rm Mpc^3}$. However, for ET this increases to $10^6\,{\rm Mpc^3}$ and for CE to $\sim3\times10^7\,{\rm Mpc^3}$. 

We will now estimate the merger rates that we can expect for these light systems, and use the distances calculated above for an SNR threshold of 12 to predict the number of events we can hope to detect per year.

\section{PBH merger rate}
\label{sec:mergerrate}

We now calculate how many PBH mergers with small mass ratios $q<10^{-2.5}$ we can hope to detect with ET and CE. We first calculate the distribution of PBHs that can be expected to form from a primordial power spectrum which is boosted by $6-7$ orders of magnitude with respect to the observed amplitude on CMB scales. We then calculate the merger rate for this population, and isolate those with mass ratios where a dark matter spike is expected to have survived around the larger of the two black holes.

We do not account for the effect of the DM spikes on the PBH merger rate, as was done in for example \cite{Kavanagh:2020cfn} for equal-mass mergers. In principle, dynamical friction from the DM spike may reduce the merger time of the binary, affecting the predicted merger rate for these systems. However, detailed calculations have not yet been performed for the large mass ratio PBH binaries we consider here (though see Ref.~\cite{Pilipenko:2022emp} for recent work on circumbinary accretion of DM). We also neglect the effects of baryonic accretion on the PBH merger rate, which is relevant only for PBH masses larger than a few solar masses~\cite{DeLuca:2020fpg}.


\subsection{PBH mass function}

As a concrete example of a  realisation of PBH formation, we assume that PBHs can form from large overdensities which collapse shortly after single-field inflation. We assume an initially Gaussian density field, and that there is no clustering, however see \cite{Kovetz,Clesse,Youngcluster,Ballesteros:2018swv,Bringmann:2018mxj,Atal:2020igj} for other possibilities. Note that PBHs can also form from e.g.\ the collapse of cosmic strings~\cite{Hawking:1987bn,Polnarev:1988dh} or during phase transitions in the early universe~\cite{Crawford:1982yz,Hawking:1982ga,Kodama:1982sf}. 

\begin{figure}[t]
     \centering
        \includegraphics[width=0.48\textwidth]{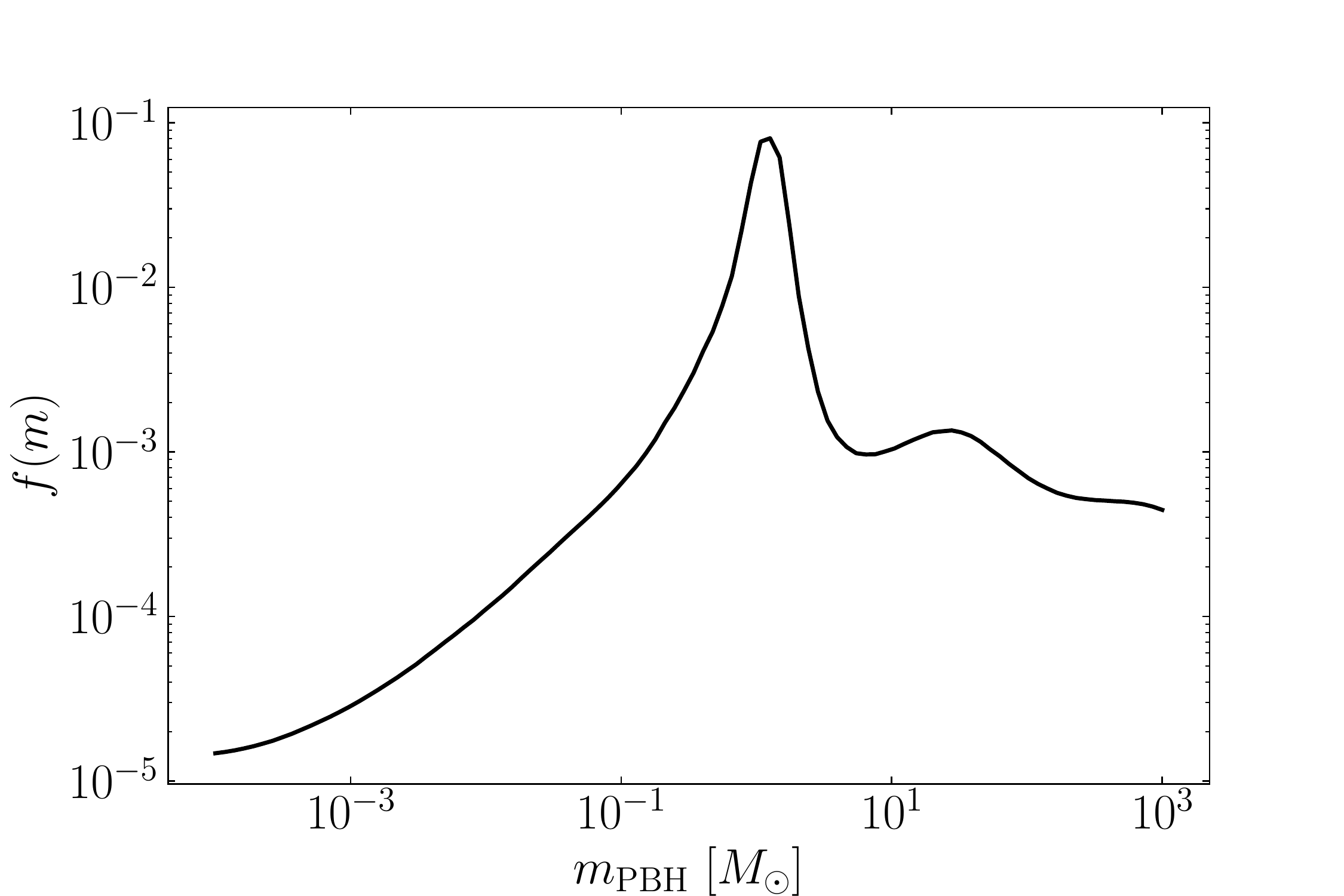}
        \caption{\textbf{The mass function of PBHs adopted in this work. } We normalize the mass function according to $\int f(m)\,\diff\ln m= \fPBH$. The multiple peak structure arises from the evolving equation of state in the early universe~\cite{CosmicConundra}. The peak of the primordial power spectrum is at $k_p=10^5\,\rm{Mpc}^{-1}$, and the abundance of PBHs in this case is $\fPBH=0.066$.}
        \label{fig:massfunction}
\end{figure}

A reference primordial power spectrum (PPS) that satisfies all current constraints can be modelled by a piece-wise power law spectrum \cite{Atal:2018neu,Byrnes:2018txb}:
\begin{align}
    \begin{split}
       \mathcal{P}_\mathcal{R}(k) &= A\begin{cases}
        \left(\frac{k}{k_p}\right)^{n_g} & k \leqslant k_p\,, \\
        \left(\frac{k}{k_p}\right)^{-n_d} & k > k_p\,.
        \end{cases} \label{eq:Pzeta_sym} 
    \end{split}
\end{align}
We choose $n_g=4$ because it is representative of peaks produced via ultra-slow-roll models of inflation \cite{Atal:2018neu,Byrnes:2018txb}, and we choose $n_d=0.05$ as being the slowest decay possible (in order to enhance production of light PBHs) without conflicting with observational constraints on scales smaller than the peak~\cite{Gow}. We then compute the mass function of PBHs that would be formed, shown in \cref{fig:massfunction}, using the Press-Schechter formalism as laid out in Ref.~\cite{Gow} and accounting for the effect of the evolving equation of state in the early universe \cite{CosmicConundra,ByrnesQCD}. This results in a mass function with multiple peaks owing to the fact that PBHs form more easily at times when the equation of state was lower. The most prominent peak is around $1\,{\rm M_\odot}$, because scales with this horizon mass were collapsing at the time of the QCD phase transition, where the equation of state decreases by a factor of $\sim1/3$~\cite{ByrnesQCD}. Since we are interested in stellar-mass black holes, we choose $k_p=10^5\,{\rm Mpc^{-1}}$ and verify that changes to $k_p$ within a factor of 10 are always dominated by the QCD effect and still preferentially produce solar-mass PBHs. This further motivates our choice of benchmark mass $m_1=1\,{\rm M_\odot}$.

We do not take into account various considerations that may affect the relationship between the amplitude of the primordial power spectrum and the PBH abundance \cite{Kalaja:2019uju}, for example non-Gaussianity, the shape of the curvature perturbation \cite{Musco:2018rwt} or the non-linear relationship between the density and curvature perturbation \cite{Young:2019yug,Kawasaki_2019,Luca_2019} because we are not trying to constrain $f_{\rm PBH}$ with a specific amplitude of the power spectrum. Instead, we fix the shape of the spectrum and then have the freedom to vary $A$ in order to obey $f_{\rm PBH}$ constraints or to absorb subtleties in the calculation of $f_{\rm PBH}$ from the initial distribution of densities. However, we note that if the initial conditions are non-Gaussian, the initial distribution of PBHs will be clustered, and this will effect the merger rate, although it is uncertain by how much as the level of non-Gaussianity increases~\cite{Ballesteros:2018swv,Young:2019gfc}.

We find that for $k_p=10^{5}\,\mathrm{Mpc^{-1}}$, an amplitude of $A=3.5\times10^{-3}$ saturates the constraints as described below in \cref{sec:PBHconstraints} and leads to $f_\mathrm{PBH}=0.066$.

\subsection{Differential merger rate calculation}

Two neighbouring PBHs may form a binary in the early universe when their self-gravity dominates over the Hubble flow~\cite{Nakamura:1997sm,Ioka:1998nz,Ali-Haimoud:2017rtz}. Two further requirements are that the radial tidal forces from all other PBHs and matter fluctuations are weaker than the attraction of the pair and tidal torques large enough to prevent a head-on collision~\cite{Raidal,Liu:2018ess}.

\begin{figure}
     \centering
        \includegraphics[width=0.48\textwidth]{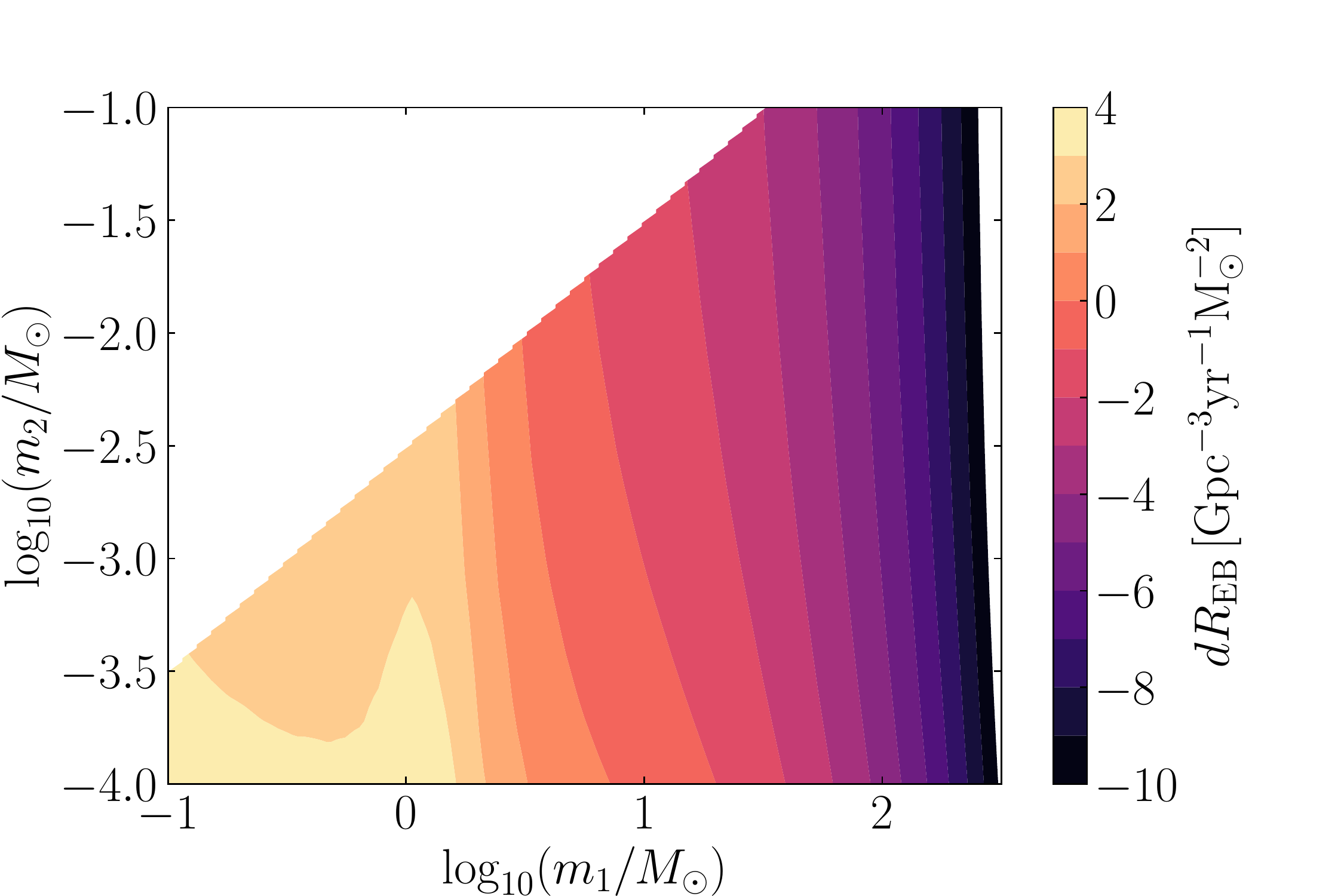}
        \caption{\textbf{Differential merger rate of binaries formed in the early universe, with $m_2<10^{-2.5}m_1$,} including suppression factor (and therefore an underestimate).}
        \label{fig:dReb}
\end{figure}

The differential merger rate for PBH binaries formed in the early universe (early-forming binaries (EB) being dominant over those formed by tidal capture in PBH clusters in the late Universe~\cite{Bird:2016dcv,Korol:2019jud,Franciolini:2022ewd}, at least for small values of $\fPBH$) with component masses $m_1$ and $m_2$ is given by~\cite{Vaskonen,Hutsi,Phukon:2021cus}:
\begin{align}
    \begin{split}
    \label{eq:dReb}
        \diff R_{EB} &= f_{\rm SUP}\frac{1.6 \times 10^6}{\mathrm{Gpc}^3 \,\mathrm{yr}} \fPBH^\frac{53}{37}\eta^{-\frac{34}{37}}\left(\frac{m_1}{M_\odot}\right)^{-\frac{32}{37}}\\ & \quad \times \left(\frac{t}{t_0}\right)^{-\frac{34}{37}}\psi(m_1)\psi(m_2)\,\diff m_1\, \diff m_2\,,
    \end{split}
\end{align}
where $f_{\rm SUP}$ is the so-called suppression factor (details below), $\eta=m_1m_2/(m_1+m_2)^2$, $t$ is the cosmic time of merger, $t_0$ is the cosmic time today, and $\psi(m)$ is the mass function defined by:
\begin{equation}
    \int m\psi(m) \,\diff \ln m = \int \frac{f(m)}{\fPBH}\, \diff\ln m =1,
\end{equation}
where the second equality demonstrates the relationship with the notation for the mass function $f(m)$ as plotted in \cref{fig:massfunction} for clarity. The differential merger rate calculated for the mass function in \cref{fig:massfunction} including the suppression factor, is shown in \cref{fig:dReb}.

However, it is very important to note that this merger rate is only reliable for $f_{\rm PBH}\ll1$, which we satisfy, because otherwise PBH clusters may form that would perturb binaries \cite{Vask2,DeLuca:2020jug,Tkachev:2020uin}. It is also only valid for relatively narrow mass functions. Since we are using a very broad mass function, and are especially reporting merger rates for low mass ratios because those are the binaries where the \gls*{dm} spike will survive, caution must be used. The reason for this is that the suppression factor assumes that a population of lower mass black holes will cause third-body disruption to binaries. However, a large population of very low mass black holes are unlikely to take part in this process very effectively. It is not clear by how much the rate is underestimated, and that study is beyond the scope of this work, so we instead report upper and lower bounds based on using $f_{\rm SUP}$ as in \cite{Phukon:2021cus}, which approximates the formulations in \cite{Vaskonen,Hutsi}
\begin{equation}
\label{eq:suppression}
    f_{\rm SUP}=S_1\times S_2\,,
\end{equation}
with
\begin{align}
    S_1&=1.42\left(\frac{\langle m_{
\rm PBH}^2\rangle/\langle m_{
\rm PBH}\rangle^2}{\overline{N}+C} + \frac{\sigma_M^2}{f_{\rm PBH}^2}\right)^{-\frac{21}{74}}e^{-
    \overline{N}}\,,\\
    S_2&\approx {
    \rm min}\left(1, 9.6\times10^{-3}f_{
    \rm PBH}^{-0.65}e^{0.03\log^2(f_{\rm PBH})}\right)\,,
\end{align}
where $\langle m_{
\rm PBH}^2\rangle$ and $\langle m_{
\rm PBH}\rangle^2$ are the variance and squared mean of the PBH mass respectively,  $\sigma_M^2=0.005$, and $\overline{N}$ (for which we use the full expression rather than the upper and lower limits given in \cite{Phukon:2021cus}), and $C$ are given by:
\begin{equation}
    \overline{N}=\frac{m_1+m_2}{\langle m_{
\rm PBH}\rangle}\frac{f_{\rm PBH}}{f_{\rm PBH}+\sigma_M}\,,
\end{equation}
\begin{align}
    \begin{split}
            C&=\frac{f_{\rm PBH}^2\langle m_{
\rm PBH}^2\rangle}{\sigma_M^2\langle m_{
\rm PBH}\rangle^2} \times\\
&\left[\left(\frac{\Gamma(\frac{29}{37})}{\sqrt{\pi}}U\left(\frac{21}{74},\frac{1}{2},\frac{5f_{\rm PBH}^2}{6\sigma_M^2}\right)\right)^{-\frac{74}{21}}-1\right]^{-1}\,,
    \end{split}
\end{align}
with $\Gamma$ the gamma function and $U$ the confluent hypergeometric function.
The expression for $\overline{N}$ determines the effect of nearby PBHs on the merger rate and using the full expression over-suppresses the merger rate. This gives us our lower bound in all of the merger rate plots, whilst $f_{\rm SUP}=1$ (i.e. no effect), gives us our upper bound. Given that we have produced the largest possible merger rate by maximising the amplitude of the PPS, a detection in the observing runs of \gls*{et} and \gls*{ce} would put tight constraints on $f_{\rm SUP}$, modulo effects on the merger rate caused by the dark matter spikes. Non-detection would constrain combinations of $f_{\rm SUP}$ and power spectrum amplitude $A$, but would not break that degeneracy.

\subsection{Saturating observational constraints}
\label{sec:PBHconstraints}

In order to determine the most optimistic merger rate in this scenario, we saturate direct constraints on the PBH abundance. We do so by increasing the amplitude of the PPS $A$ until we hit the strongest constraints. At the time of writing, LIGO/Virgo stochastic gravitational wave background (SGWB) spectrum constraints due to BBH mergers\footnote{Constraints from the direct observations of individual mergers with LIGO/Virgo/Kagra are typically weaker than those from the SGWB in the sub-solar mass range (see e.g.~\cite{Gow:2019pok,DeLuca:2020qqa,Hall:2020daa,Garcia-Bellido:2020pwq,Wong:2020yig,DeLuca:2021wjr,Franciolini:2021tla,Chen:2021nxo,Hutsi}).} and the envelope of microlensing constraints from various probes on the sub-solar mass range are the strongest on the relevant scales~\cite{Green:2020jor,PBHbounds}. 

We calculate the resulting stochastic gravitational wave signal due to all mergers by inputting \cref{eq:dReb} into the expression from \cite{Phinney:2001di,Zhu:2011bd,Vaskonen}
\begin{align}
\begin{split}
\frac{\diff \Omega_{\rm GW}(f)}{\diff m_1\, \diff m_2\,\diff z} &= \frac{f}{\rho_cc^2H(z)(1+z)}\frac{\diff E_{GW}}{\diff f}(f_r)\\
&\qquad\qquad \times \frac{\diff R_{\rm EB}(m_1,m_2)}{\diff m_1\, \diff m_2}\,,
\end{split}
\end{align}
with $H(z)$ the Hubble factor and $\rho_c$ the critical density. The observed GW frequency $f$ is related to the source-frame frequency as $f_r = (1+z)f$, and the GW spectrum $\diff E_\mathrm{GW}/\diff f$ from merging BHs  is given in App.~\ref{app:GWspectrum}.

We then integrate over all $m_1$ and $m_2$, dividing by 2 to avoid double-counting, and we integrate between $z=0$ and $z_{\rm eq}$. 
We increase the amplitude of the primordial power spectrum (which filters through to $f(m)$) until the spectrum without suppression factor hits the LIGO/Virgo O5 sensitivity curve. This means that the merger rates we now report will be viable by the time ET and CE come online in terms of being compatible with other gravitational wave constraints that will be realised in the meantime.

The microlensing constraints do not depend directly on the merger rate, since they search for distinctive variatons in stellar light curves due to individual black holes passing in front of stars in the centre of the Milky Way or in nearby galaxies~\cite{Macho:2000nvd,EROS-2:2006ryy,Niikura:2017zjd,Niikura:2019kqi}.
Therefore, using our results for the mass function $f(m)$ and the prescription in Ref.~\cite{extmass}, we check that the values of $f_{\rm PBH}$ corresponding to $k_{\rm p}=10^5\,{\rm Mpc^{-1}}$ and $A=3.5\times10^{-3}$ which saturates the SGWB constraints is below the threshold allowed by the microlensing constraints from figure 3 of \cite{BradAnne}. Indeed, $f_{\rm PBH}=0.066$ whilst the constraints require $f_{\rm PBH} < 0.78$. 

\subsection{Binned merger rates}

\begin{figure}
     \centering
         \includegraphics[width=0.48\textwidth]{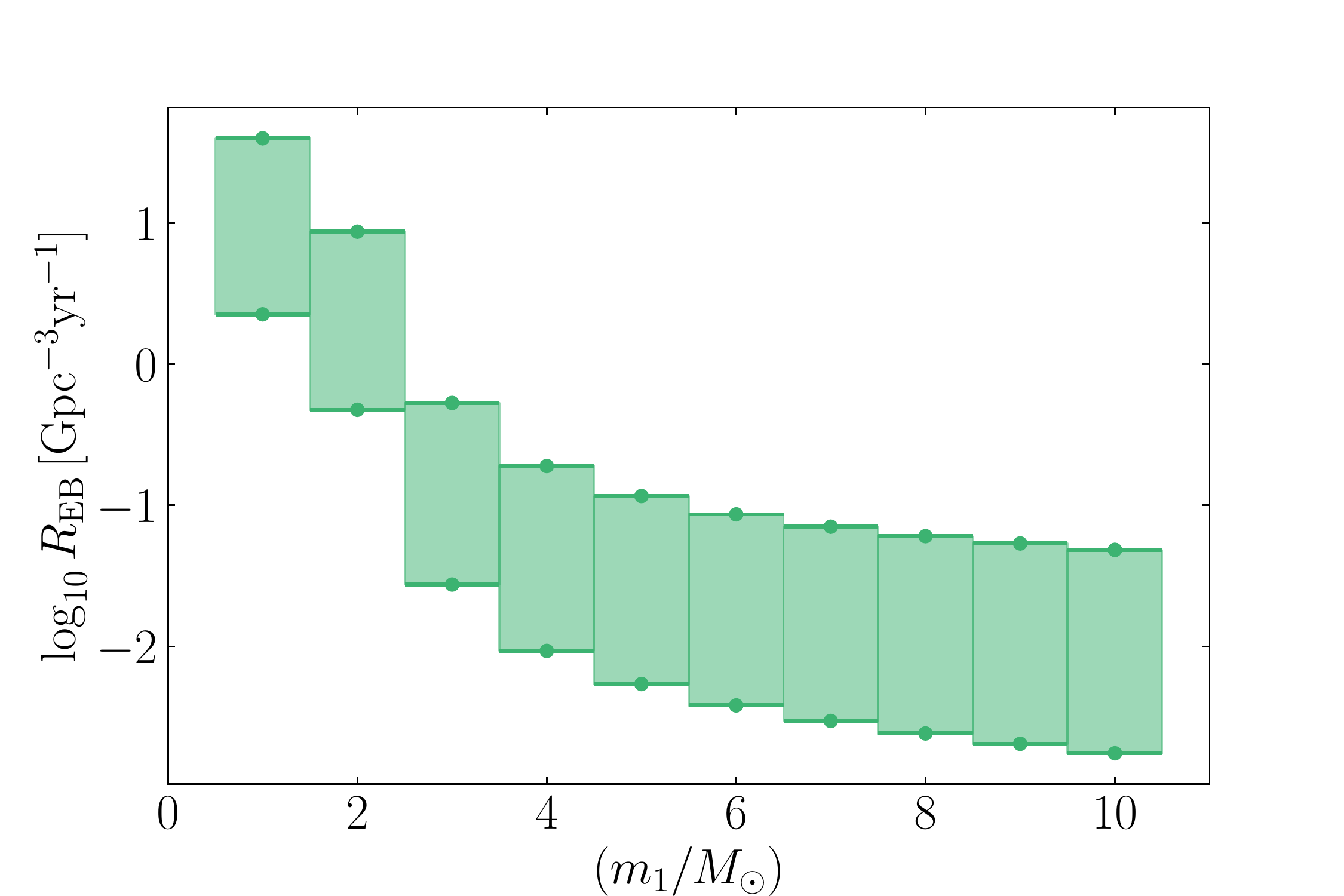}

        \caption{\textbf{PBH Merger rate for $m_1\pm\,0.5\,M_\odot$, including mergers with all secondary masses below $m_2 <10^{-2.5}\,m_1$}. The lower value of the filled region is calculated including the suppression factor of \cref{eq:suppression} (and therefore an underestimate), whilst the upper value is calculated without. All rates are averaged between redshift 0 and 0.1.}
        \label{fig:Reb}
\end{figure}

The merger rates of a central black hole of mass $m_1$, within a bin of width $\pm 0.5\,{\rm M_\odot}$, merging with all masses below $10^{-2.5}\,m_1$ are shown in \cref{fig:Reb}. All rates are averaged between redshifts 0 and 0.1, and the upper and lower limits of the filled regions are calculated without and with the suppression factor respectively. 
The maximum merger rate we find is $R_{\rm EB}\sim40\,{\rm Gpc^{-3}yr^{-1}}$ for $m_1=1\,{\rm M_\odot}$ without suppression factor, and $R_{\rm EB}\sim2\,{\rm Gpc^{-3}yr^{-1}}$ with the suppression factor which acts as a lower bound.

\begin{figure}
    \centering
    \includegraphics[width=0.48\textwidth]{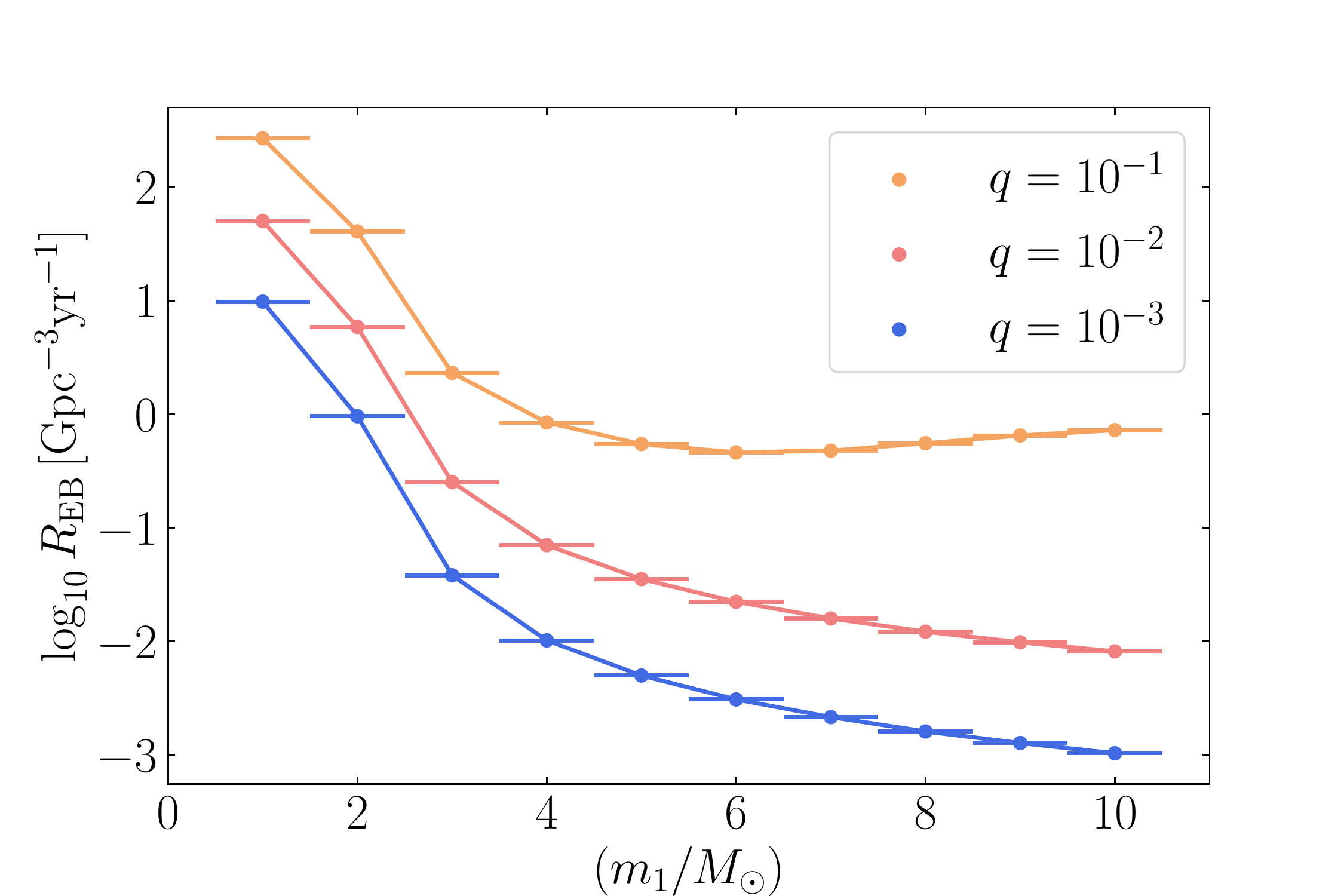}
    \caption{\textbf{Merger rate of black hole binaries with $q=10^{-3},\,10^{-2},\,10^{-1}$}, without suppression factor. All averaged between redshift 0 and 0.1. The merger rate for our benchmark system is given by the left-most bin of the $q=10^{-3}$ case.}
    \label{fig:Reb_comb}
\end{figure}

In order to report an event rate for a specific benchmark system, we can combine the binned merger rate for $m_1=1\pm0.5\,{\rm M_\odot}$ and $m_2=10^{-3}\,m_1$, shown in figure \cref{fig:Reb_comb}, with the observable volumes of each detector from \cref{sec:reach}. We find that aLIGO, ET and CE can expect to observe $\num{2e-6}$, $\num{8e-3}$ and 0.3 events with an SNR of at least 12 per year respectively.

Furthermore, we would expect to see larger mass ratio mergers (where the spike will have been disrupted) occurring in vacuum. For example, for mass ratios of $q=10^{-1}$ and $q=10^{-2}$, the expected merger rates for our most optimistic model are also shown in \cref{fig:Reb_comb}. LIGO/Virgo searches such as Refs.~\cite{Nitz1,Nitz2} can probe these mass ranges but are not sensitive enough to probe these merger rates yet. However, ET and CE will be sensitive enough to search for these vacuum systems \cite{Barsanti:2021ydd}. If the merger rate for this region of the parameter space was observed to be in line with the rates presented in \cref{fig:Reb_comb}, this would be a strong hint that the more extreme-mass ratio systems, with dark matter spikes, exist.

\subsection{Spike survival}
\begin{figure}
    \centering
    \includegraphics[width=0.48\textwidth]{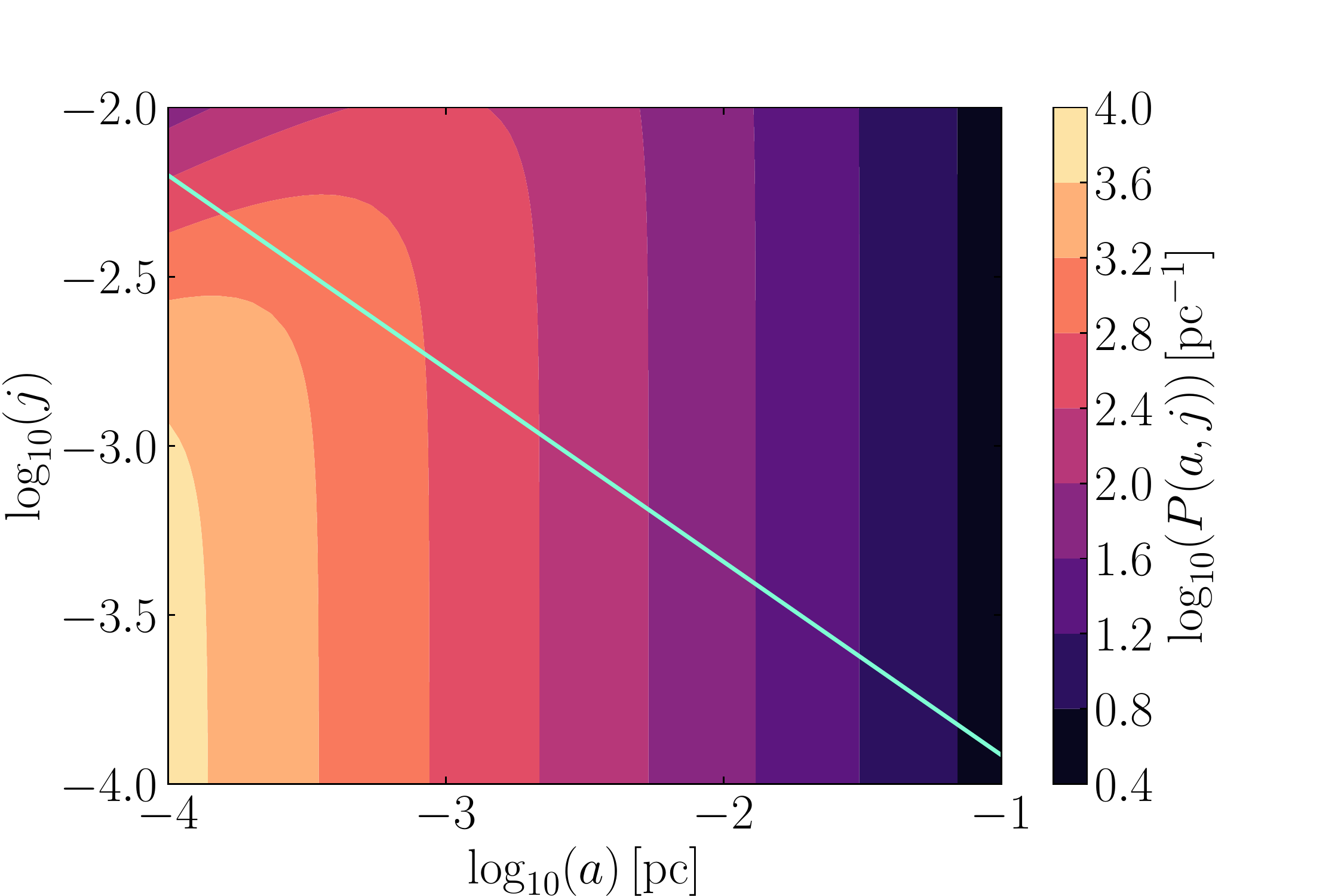}
    \caption{\textbf{Probability distribution of PBH binaries that decouple in the early universe as a function of semi-major axis and angular momentum.} The green line represents a coalescence time of 13 Gyr.}
    \label{fig:Pja}
\end{figure}

Finally, in order to ensure that the spike will survive the first few encounters of the inspiral, we check that the radius of closest approach is larger than the separation of the binary at the lowest value of the detectors' frequency range, $f_{\rm low}$, for our benchmark masses. The radius of closest approach is 
\begin{equation}
    r_{\rm min} = a(1 - \sqrt{1 - j^2}),
\end{equation}
where $a$ is the semi-major axis of the binary and $j$ is the angular momentum (related to the eccentricity by $j = \sqrt{1-e^2}$). We fix the time of coalescence of the binary system to be $13\,{\rm Gyr}$,
\begin{equation}
    t_{\rm coa} = \frac{3a^4j^7c^5}{85m_1m_2(m_1+m_2)G^3} = 13\,{\rm Gyr}
\end{equation}
and then select the most probable values of $a$ and $j$ that satisfy this using equation (5) of \cite{BVKmerge} to compare with $r_{\rm low}= (G(m_1+m_2)/(f_{\rm low}^2\pi^2))^\frac{1}{3}$. We plot the probability distribution in \cref{fig:Pja}, highlighting $t_{\rm coa}=13\,{\rm Gyr}$ with the green line. For a relevant range of values of $a$ and $j$, we find $r_{\rm min}/r_{\rm low}>100$. This means that the first encounter while the orbit is still elliptical should not destroy the spike within the radius that corresponds to the lower end of the frequency band of \gls*{et} and \gls*{ce}, $f_{\rm low}=1\,{\rm Hz}$.

Then, using the definition of the tidal radius of the companion object from equation (4) of \cite{vdBosch} which is a refinement of the Roche limit 
that does not treat the central black hole as a point mass, we more stringently check that for a separation given by the radius of closest approach $r_{\rm min}$, the tidal radius is still greater than $r_{\rm low}$. This is indeed the case with $r_{\rm tidal}/r_{\rm low}>10$ for the most probable range of semi-major axes.

\section{Dephasing due to dark matter spike}
\label{sec:dephasing}

Now that we have seen that there could be a number of detectable events per year, we turn our focus to how we will need to go about detecting them.


We have seen in the previous section that PBHs cannot explain the entire DM budget in the mass range relevant for ET and CE searches. However, if they account for a sub-dominant fraction of the DM, they must be accompanied by a component of particle DM making up the remainder. Both analytical calculations and verifying simulations~\cite{Bertschinger:1985pd,Mack:2006gz,Gosenca,Boudaud:2021irr} have shown that in this scenario, (cold, collisionless) DM particles will form dense spikes around PBHs with a distinct power law density profile
\begin{equation}
\label{eq:pbhspike}
    \rho(r)=\rho_\mathrm{sp}\left(\frac{r_\mathrm{sp}}{r}\right)^{\gamma_s}\,,
\end{equation}
where $\gamma_s=9/4$, $\rho_{\rm sp}=\frac{1}{2}\rho_{\rm eq}$ and $r_{\rm sp}=\left(2Gm_1t_{\rm eq}^2\right)^{1/3}$~\cite{Gosenca}. Here, $r$ is the radial distance from the central PBH of mass $m_1$, with $\rho_{\rm eq}$ and $t_{\rm eq}$ being the background density and cosmic time at matter-radiation equality. We neglect relativistic corrections to the simple power-law which occur for radii $r \sim r_\mathrm{isco}$~\cite{Sadeghian:2013laa,Ferrer:2017xwm,Speeney:2022ryg}. At these small radii, GW emission (rather than dynamical friction) is expected to dominate. A self-consistent description of dynamical friction and feedback in the relativistic regime is not yet available, though see Ref.~\cite{Speeney:2022ryg} for estimates of the impact of post-Newtonian corrections in these systems.

It is useful to reparametrize the density profile in terms of the DM density $\rho_6$ at a distance $r_6 \equiv 10^{-6}\,\mathrm{pc}$ from the BH, as in Ref.~\cite{Coogan:2021uqv}:
\begin{equation}
\label{eq:pbhspike_rho6}
    \rho(r)=\rho_6\left(\frac{r_6}{r}\right)^{\gamma_s}\,.
\end{equation}
For the purposes of parameter estimation, we treat $\rho_6$ as a free parameter which controls the overall normalization of the spike density.
Of course, for a specific formation scenario, the benchmark value of $\rho_6$ will depend on $m_1$; equating \Cref{eq:pbhspike} and \Cref{eq:pbhspike_rho6}, we see that for a PBH mass $m_1$, the benchmark density normalization should be $\rho_6 = 1.396 \times 10^{13}\, (m_1/M_\odot)^{3/4} \,M_\odot/\mathrm{pc}^3$.\footnote{This value of $\rho_6$ differs slightly from that given for PBHs in Ref.~\cite{Coogan:2021uqv}, as we here use slightly updated cosmological parameters. In this work, we take $\rho_\mathrm{eq} = 2605 \,M_\odot/\mathrm{pc}^3$ and $t_\mathrm{eq} = 1.617\times 10^{12}\,\mathrm{s}$, calculated using \texttt{CAMB}~\cite{Lewis:1999bs,2012JCAP...04..027H}, assuming Planck-2018 cosmology~\cite{Planck:2018vyg}.}

If the dominant DM component is made up of canonical WIMP-like DM, then the dense DM spikes around PBHs would give rise to a large gamma-ray flux due to WIMP annihilation. Constraints from point source searches and the diffuse gamma-ray flux therefore severely limit the possibility of mixed PBH-WIMP DM scenarios~\cite{Lacki:2010zf,Gosenca,Bertone:2019vsk}. However, if the remaining DM is composed of another candidate, such as axion-like particles or asymmetric WIMPs, these constraints can be evaded and these dense DM spikes may have an impact on GW searches for PBHs~\cite{Brad_pbh_merge}.

Gravitational wave searches in the LIGO/Virgo data have predominantly used matched filtering, where a bank of gravitational waveform templates which are calculated for a particular set of parameters are compared to the data to search for a `match' with a pre-specified signal-to-noise threshold. Thus far, these gravitational waveforms have been modelled assuming that the inspiral and merger have occurred in vacuum. However, the gravitational waveform looks different if the inspiral instead occurs in non-empty space, namely a DM spike.
We must, therefore, use non-vacuum waveforms to avoid missing signals in the data due to SNR loss, and to avoid mischaracterizing signals, where the use of vacuum waveforms may lead to  biased parameter reconstruction.


As the smaller BH moves through the spike of DM particles, they will form a wake which imparts a drag force on the BH, reducing its orbital velocity. This in turn causes the smaller BH to drop into a lower orbit more quickly than it would in vacuum. This effect is known as dynamical friction~\cite{Chandrasekhar1943a,Chandrasekhar1943b,Chandrasekhar1943c}. The inspiral happens in fewer GW cycles, causing the gravitational waveform to gradually go out of phase with respect to the vacuum case, an effect known as `dephasing'. Full details of the state-of-the-art in calculating the DM dephasing effect are given in Ref.~\cite{Kavanagh:2020cfn,Coogan:2021uqv}; here we briefly summarize the physics and our numerical approach.

Working at Newtonian order, the evolution of the binary separation $r$ for quasi-circular orbits can be described by:
\begin{equation}
\begin{split}
    &\dot{r} = - \frac{64\, G^3\, M \, m_1\, m_2}{5\, c^5\, r^3} \\ &- \frac{8 \pi\, G^{1/2}\, m_2 \,   r^{5/2} \, \rho_\rmDM(r,t) \,\xi(r, t) \,\log\Lambda}{\sqrt{M} m_1 }  \, ,
    \label{eq:r_eom}
\end{split}    
\end{equation}
where $M = m_1 + m_2$ is the total mass of the binary. The first term on the right hand side corresponds to GW emission, while the second corresponds to dynamical friction.\footnote{We neglect contributions from accretion onto the smaller BH and the varying enclosed mass due to the DM spike; in the absence of feedback, these effects are expected to be sub-dominant~\cite{Macedo:2013qea,Cardoso:2019rou}.} The dynamical friction force traces the density profile of the DM within the spike $\rho_\mathrm{DM}(r,t)$, while the factor $\xi(r, t)$ corresponds to the fraction of DM particles at a given radius $r$ moving more slowly that the local orbital speed, which are those relevant for the calculation of the dynamical friction force~\cite[Sec.~8.1]{BinneyAndTremaine}. The Coulomb logarithm $\log\Lambda$ incorporates information about the range of distances from the smaller BH at which gravitational scattering with DM particles is effective. Following Ref.~\cite{Kavanagh:2020cfn}, we take $\Lambda = \sqrt{m_1/m_2}$. 

During the inspiral, the motion of the BH binary will inject energy into the DM spike, altering its density profile. In Ref.~\cite{Kavanagh:2020cfn}, the DM spike was described in terms of a spherically symmetric, isotropic distribution function, which is altered by the gravitational scattering of DM particles with the orbiting BH. Following the evolution of this distribution allows us to calculate the time evolution of the DM density and velocity distribution, $\rho_\mathrm{DM}(r,t) \xi(r, t)$, while simultaneously solving for the binary separation $r$ through \Cref{eq:r_eom}.  Given a trajectory $r(t)$ the GW frequency is calculated as:
\begin{equation}
\label{eq:fKep}
    f(t)=\frac{1}{\pi}\sqrt{\frac{GM}{r(t)^3}} \,,
\end{equation}
from which the strain and phase to coalescence can be calculated using \cref{eq:strain_and_phase}.

The feedback formalism described above was implemented in the publicly-available \texttt{HaloFeedback} code~\cite{HaloFeedback}. The most prominent effect is that the spike will be locally depleted as some particles are ejected, reducing the size of the dephasing effect. Taking into account this effect requires time-consuming numerical simulations using \texttt{HaloFeedback}. However, for the parameter estimation that we would like to conduct, we need to produce waveforms for a densely sampled region of the parameter space, and therefore we will rely on the analytic model that was proposed and used in \cite{Coogan:2021uqv} to describe the dephasing including feedback. In this model, the phase to coalescence $\Phi(f)$ approximately follows a broken power-law, with the break frequency $f_b$ a function of $m_1, m_2$ and $\gamma_s$. The functional form for $f_b(m_1, m_2, \gamma_s)$ was fit to \texttt{HaloFeedback} simulations of binaries with total masses in the range $M \sim 10^{3} - 10^{4}\, M_\odot$. However, as we show in \Cref{sec:HalofeedbackValidation}, this parametrization also provides an accurate fit to the behaviour of the much lighter systems we consider here. We therefore rely on this parametrization to calculate the waveforms of light, dressed PBH binaries in the following sections.

\section{Assessing detectability, discoverability and measurability}
\label{sec:measurability}

In order to assess the prospects for concretely measuring DM spikes around PBHs with ground-based GW observatories, we follow the approach of Ref.~\cite{Coogan:2021uqv}. Assuming Gaussian noise, the likelihood can be written as
\begin{equation}
    p\qty( d | h_{\bm{\theta}} ) \propto \exp\qty[ \braket{h_{\bm{\theta}}}{d}^2 - \frac{1}{2} \braket{h_{\bm{\theta}}} ] \, ,
    \label{eq:likelihood}
\end{equation}
where $d(t)$ is the strain time series data measured by the detector (which we assume to match the our benchmark signal $s(t)$)  and $h_{\bm{\theta}}$ is a model waveform $h_{\bm{\theta}}(t)$ with parameters $\bm{\theta}$. In \Cref{eq:likelihood}, the noise-weighted inner product is defined as:
\begin{equation}
    \braket{a}{b} = 4 \Re \int_0^\infty \dd{f} \frac{\tilde{a}(f)^* \, \tilde{b}(f)}{S_n(f)} \, , \label{eq:inner-product}
\end{equation}
where $S_n(f)$ is the noise power spectral density  (effectively the sensitivity curve of a given detector as a function of GW frequency). 
For Einstein Telescope, we assume the ``ET-D" configuration~\cite{Hild:2010id}, making use of  sensitivity estimates provided by the ET collaboration\footnote{\url{http://www.et-gw.eu/index.php/etsensitivities}}. For Cosmic Explorer, we assume the sensitivity as given in Ref.~\cite{CENoisePSD}. In all cases, we adopt the sensitivity averaged over sky locations, polarizations and binary orientations.

The parameters $\bm{\theta} = \bm{\theta}_\mathrm{int} \cup \bm{\theta}_\mathrm{ext}$ which describe the model waveform $h_{\bm{\theta}}$ can be divided into intrinsic parameters $\bm{\theta}_\mathrm{int}$, which describe the properties of the source, and extrinsic parameters $\bm{\theta}_\mathrm{ext}$, which depend on the observer. The intrinsic parameters describing the systems are:
\begin{align}
    \text{Vacuum:} \quad \bm{\theta}_{V,\mathrm{int}} &= \{ \mathcal{M} \}\\
    \text{Dark Dress:} \quad \bm{\theta}_{D,\mathrm{int}} &=\{ \gamma_s, \rho_6, \mathcal{M}, \log_{10} q \}\,,
\end{align}
where $q = m_2/m_1$ is the mass ratio of the binary.\footnote{We define the masses in the detector frame $m_\mathrm{det}$ related to the source-frame mass $m_\mathrm{src}$ by $m_{\mathrm{det}}=m_{\mathrm{src}}(1+z)$. Note that at a luminosity distance of $d_\mathrm{L} \sim 200 \,\mathrm{Mpc}$ (roughly the detectability horizon for these systems with ET and CE, as shown in \cref{fig:snrs}), the redshift is $z \sim 0.05$, so we expect only a small correction distinguishing between detector- and source-frame masses.} Assuming the detector's response is constant over the duration of the waveform, the extrinsic parameters $\bm{\theta}_\mathrm{ext}$ we consider are the luminosity distance to the system and the phase and time at coalescence:
\begin{equation}
    \bm{\theta}_\mathrm{ext} \equiv \qty{ d_L, \phi_c, \tilde{t}_c } \, . \label{eq:theta-ext}
\end{equation}
In practice, at each point in parameter space, we maximise the likelihood over the extrinsic parameters using a fast Fourier transform, as described in Ref.~\cite{Coogan:2021uqv}. This maximised likelihood is denoted $p_\mathrm{max}(d|h_{\bm{\theta}})$.

To compute the noise-weighted inner product, we integrate over the frequency range beginning a week before the system being observed coalesces and ending at the ISCO frequency. The use of this fixed frequency window introduces a subtle issue: the waveform being compared with the observation may last for more or less than a week over this frequency window. This means that it is not quite correct to maximize the noise-weighted inner product using a Fourier transform. Fixing this issue requires either numerically maximizing over $\tilde{t}_c$ or including it in the parameter estimation. We postpone further investigation to future work.

We explore the posterior distribution $p(\bm{\theta}|d) \propto p_\mathrm{max}(d|h_{\bm{\theta}}) p(\bm{\theta})$ using nested sampling~\cite{Skilling2004,2019S&C....29..891H}, implemented in the code \texttt{dynesty}~\cite{dynesty}. The priors we take for the intrinsic parameters are summarized in Table~\ref{tab:priors}. The initial slope of the density profile around PBHs has been analytically predicted and numerically verified (e.g.~\cite{Gosenca}) as $\gamma_s=2.25$, hence the reasonably narrow prior on this parameter, given that for these masses, we must be observing primordial black holes. We allow for a wide prior on the density normalisation which includes the vacuum value, and a range of mass ratios that have an upper bound of $10^{-2.5}$ where the assumptions on the survival of the spike break down. Mapping out the posteriors allows us to assess the question of \textit{measurability}: how well the properties of binary with a DM dress can be measured or constrained.

\begin{table}[tb!]
    \centering
    \begin{tabular}{c c}
        \toprule
        Parameter &  Prior range \\
        \colrule
        $\gamma_s$ & $[2.0, 2.5]$ \\
        $\rho_6$ [$10^{15}$ \si{\solarmass/\parsec^3}] &  $[0, 2]$ \\
        $\mathcal{M}$ [\si{\solarmass}]  & -- \\
        $\log_{10} q$ & $[-4.5, -2.5]$ \\
        \botrule
    \end{tabular}
    \caption{\textbf{Prior ranges on intrinsic parameters.} In each case, we assume a uniform prior over the range given above. Since we use the same wide prior on the chirp mass for the dark dress and vacuum systems, its precise value cancels out in the Bayes factor and does not matter.}
    \label{tab:priors}
\end{table}

In order to address the question of \textit{discoverability} (i.e.\ whether a given dark dress waveform can be distinguished from a GR-in-vacuum waveform), we compute the Bayes factor for the dark dress and vacuum models, defined as the evidence ratio:
\begin{equation}
    \operatorname{BF}(d) \equiv \frac{p\qty(d | \mathrm{D})}{p\qty(d | \mathrm{V})} \, . \label{eq:bayes-fact}
\end{equation}
The evidence for a given model is defined as:
\begin{equation}
    p\qty( d ) = \int \dd{\bm{\theta}} p_\mathrm{max}\qty(d | h_{\bm{\theta}}) \, p(\bm{\theta}) \,, \label{eq:evidence}
\end{equation}
where $\bm{\theta} = \bm{\theta}_{D}$ or $\bm{\theta} = \bm{\theta}_{V}$ for the dark dress and vacuum models respectively. Large values of the Bayes Factor ($\operatorname{BF} > 100$) correspond to strong evidence in favor of a dark dress system, compared to a GR-in-vacuum system~\cite{jeffreys1998theory,10.2307/2291091}.

\begin{figure}
    \centering
    \includegraphics[width=\linewidth]{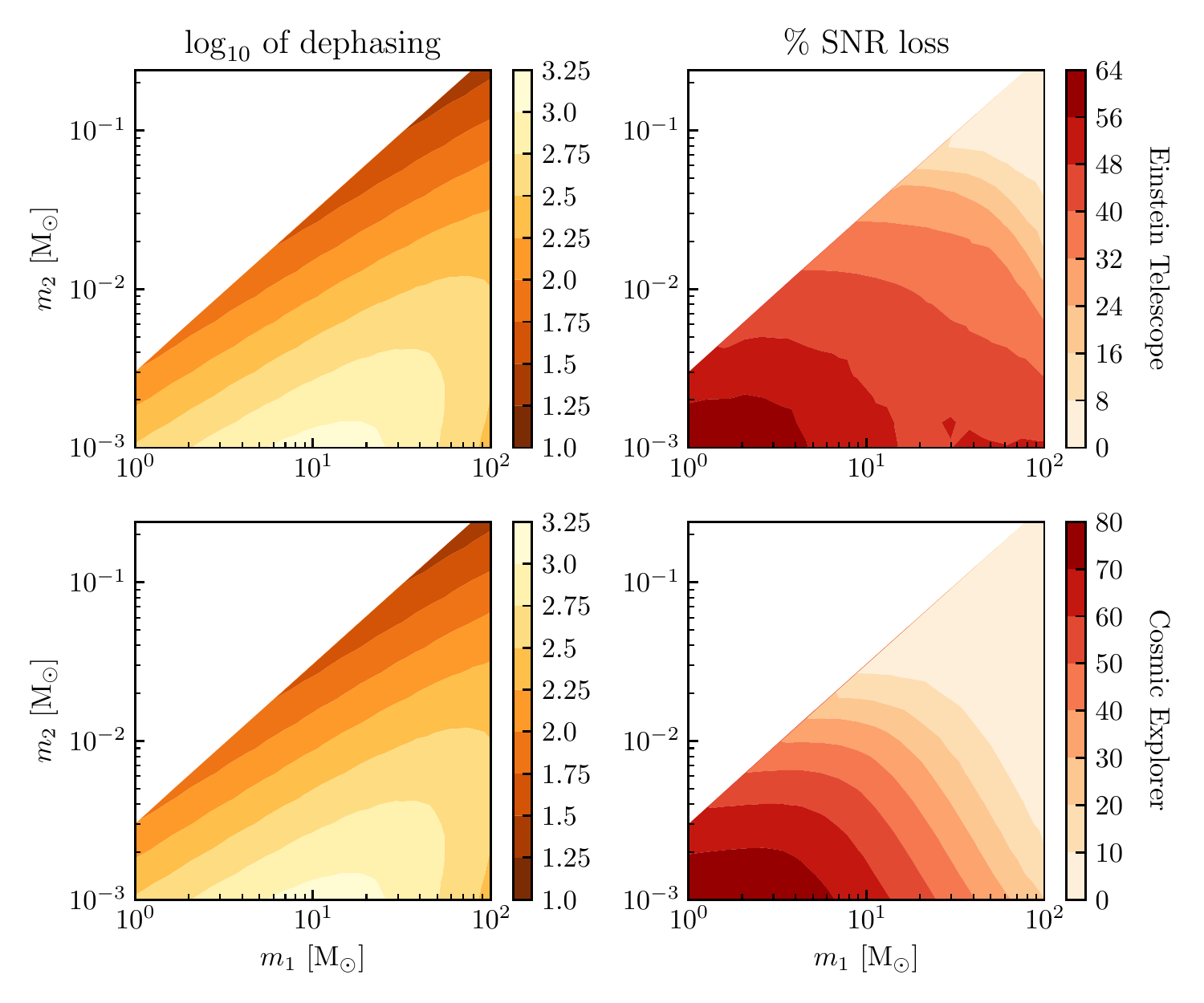}
    \caption{\textbf{Dephasing and SNR loss between best-fit vacuum template and dark dresses with various masses.} The rows correspond to ET and CE. Assumes one week of observing with the system at a distance such that it is detected with an SNR of 12 by each detector.}
    \label{fig:discoverability}
\end{figure}

\begin{figure*}
    \centering
    \includegraphics[width=\textwidth]{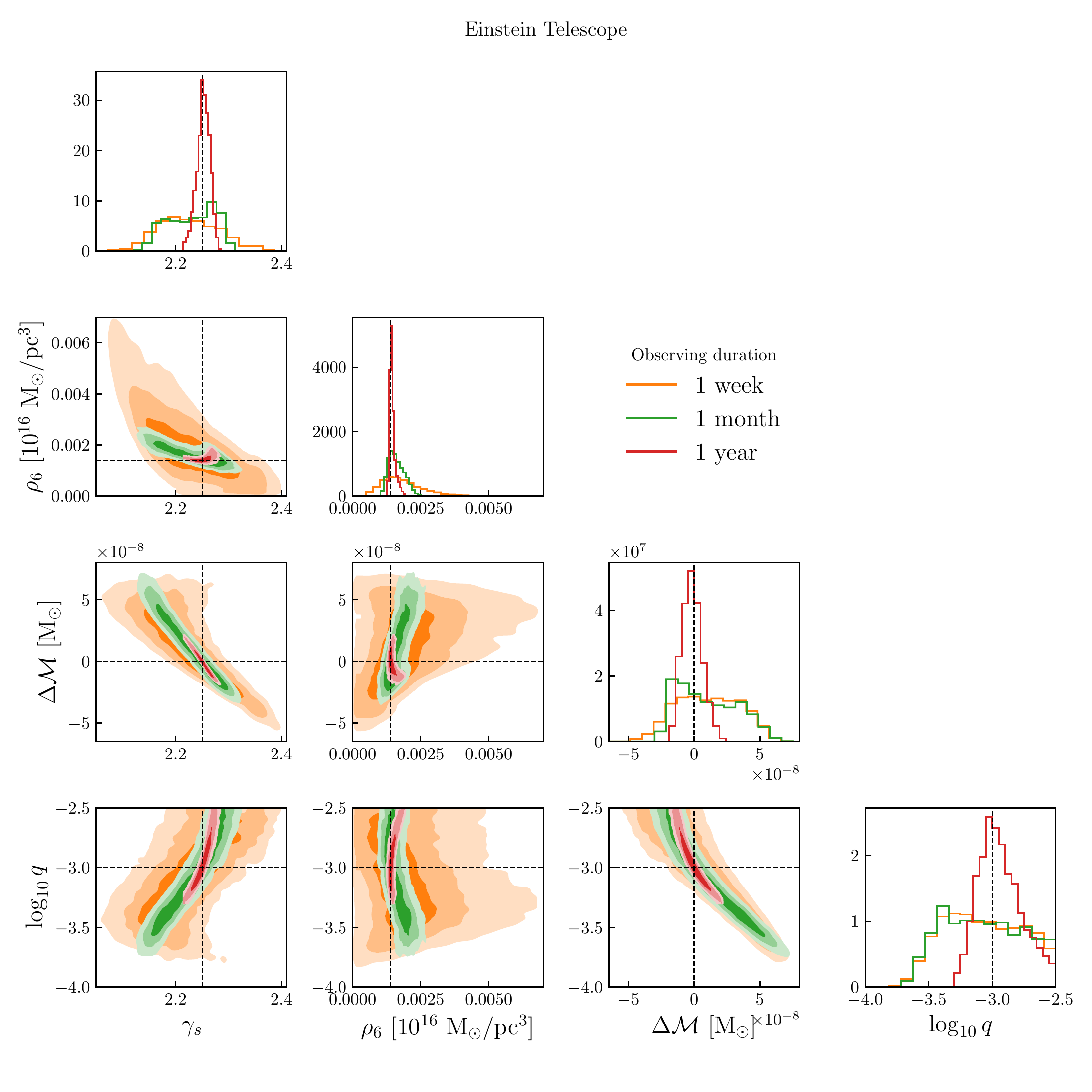}
    \caption{\textbf{One- and two-dimensional marginal posteriors for the intrinsic parameters of PBH dark dresses observed with Einstein Telescope for one week (orange), one month (green) or one year (red).} The dashed black lines indicate the true parameter values. The black hole masses are set to the benchmark values of \SI{1}{\solarmass} and \SI{e-3}{\solarmass}, and $\Delta \mathcal{M}$ is the difference between the measured and true chirp mass ($\sim \SI{1.585e-2}{\solarmass}$). The distances for the systems are set so each is detected with an \gls*{snr} of 12. The shaded contours show the \SI{65}{\percent}, \SI{95}{\percent} and \SI{99.7}{\percent} credible regions.}
    \label{fig:post-et-week-month-year}
\end{figure*}

\begin{figure*}
    \centering
    \includegraphics[width=\textwidth]{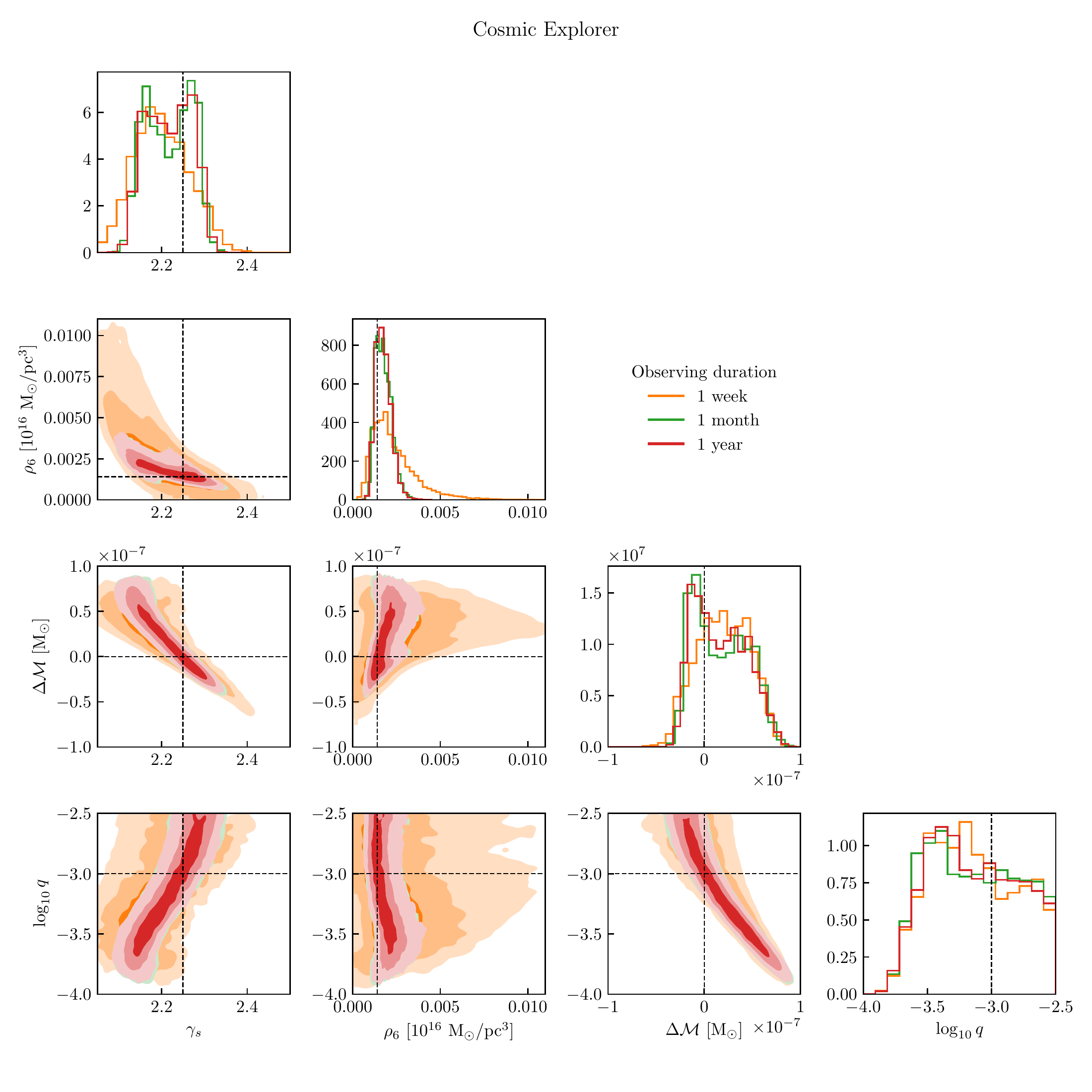}
    \caption{\textbf{Marginal posteriors for a PBH dark dress as in \Cref{fig:post-et-week-month-year}, but instead observed with Cosmic Explorer.} The system's distance was set to give an \gls*{snr} of 12 at \gls*{ce}.}
    \label{fig:post-ce-week-month-year}
\end{figure*}

\begin{figure*}
    \centering
    \includegraphics[width=0.9\textwidth]{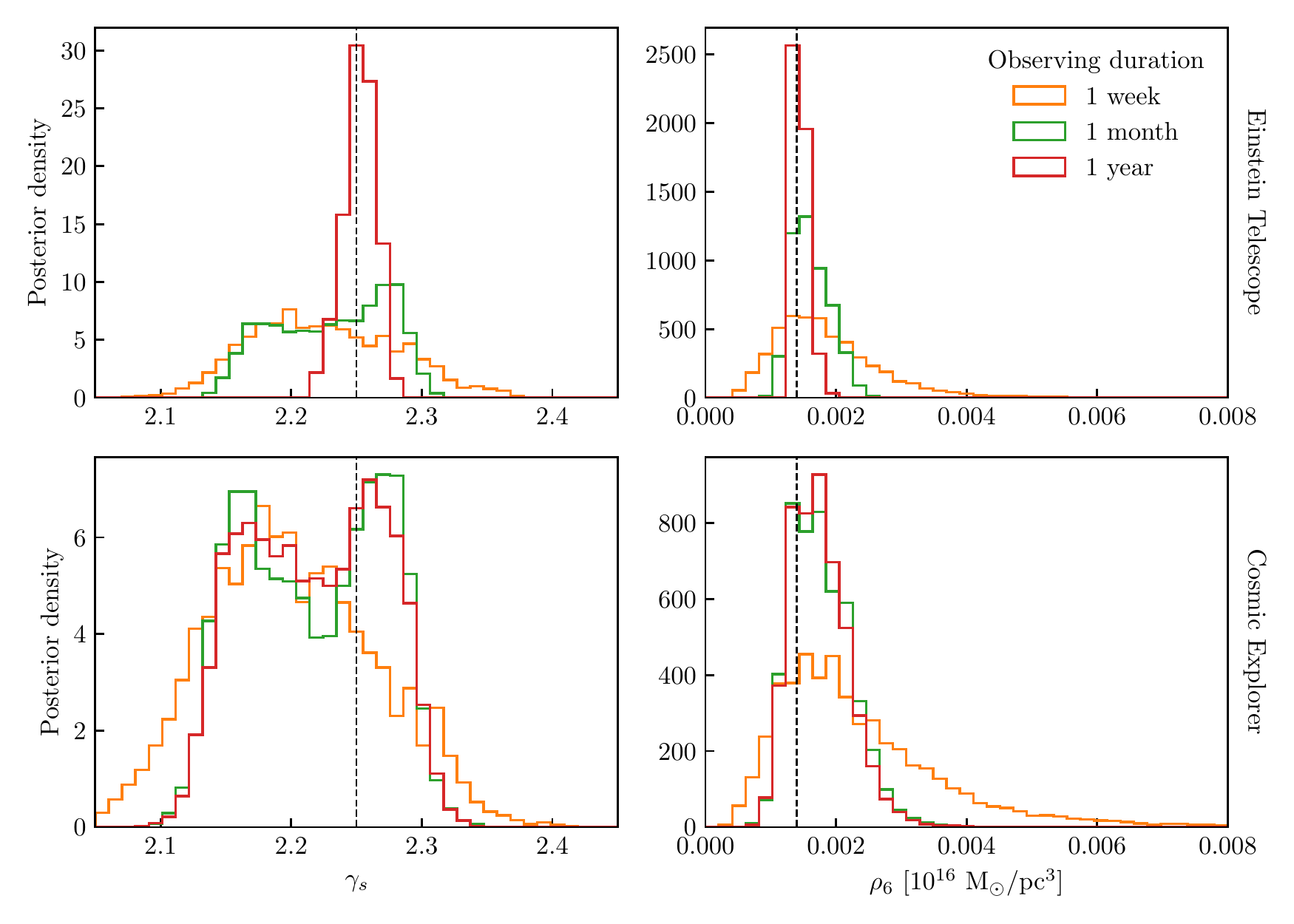}
    \caption{\textbf{The 1D marginal posteriors for $\rho_6$ and $\gamma_s$ obtained using Einstein Telescope and Cosmic Explorer.} Note the target dark dress systems in the two rows are \emph{not} the same. Their distances differ since each is observed with an \gls*{snr} of 12 in its respective detector.}
    \label{fig:post-rho6-gammas}
\end{figure*}

\section{Results}
\label{sec:results}

\begin{table}
    \centering
    \begin{tabular}{c c c}
        \toprule
        Parameter & Einstein Telescope & Cosmic Explorer \\
        \colrule
        $\gamma_s$ & $2.22_{-0.05}^{+0.07}$ & $2.20_{-0.06}^{+0.07}$ \\
        $\rho_6$ [$10^{13}$ \si{\solarmass/\parsec^3}] & $1.7_{-0.6}^{+0.8}$ & $2.0_{-0.8}^{+1}$ \\
        $\mathcal{M}$ [\si{\solarmass}] & $\num{1584577}_{-3}^{+3} \times 10^{-8}$ & $\num{1584578}_{-3}^{+3} \times 10^{-8}$ \\
        $\log_{10}(q)$ & $-3.1_{-0.3}^{+0.4}$ & $-3.2_{-0.3}^{0.4}$ \\
        \botrule
    \end{tabular}
    \caption{\textbf{Parameters inferred for benchmark system with one week of observations by different detectors.} The error bars indicate the 68\% credible intervals.}
    \label{tab:measurements}
\end{table}

\paragraph{Detectability.}
We find that aLIGO, ET and CE will be able to detect our benchmark system with a SNR of 12 out to a distance of 6.5, 78 and $\SI{286}{\mega\parsec}$ respectively (see \cref{fig:snrs}). These distances together with the upper bound on the (binned) merger rate for systems with $m_1=\SI{1\pm0.5}{\solarmass}$ and $m_2=10^{-3}\,m_1$ corresponds to event rates of $\num{2e-6}$, $\num{8e-3}$ and 0.3 per year. Increasing the mass of the systems would increase the distance out to which ET and CE will be able to detect the system, but the price to pay is that the merger rate decreases with larger $m_1$ as shown in \cref{fig:Reb_comb}.

\medskip

\paragraph{Discoverability.} We have evaluated the Bayes factor for a PBH binary with $(m_1, m_2) = (10, 10^{-2}) \,\SI{}{\solarmass}$, assuming 1 week of observation with ET and CE. We find $\mathrm{BF} \approx 10^{18}$ for ET and $\mathrm{BF} \approx \num{6e6}$ for CE, indicating overwhelming evidence in favor of the presence of a dark dress in the system in both cases. For this system we also find that the dephasing with respect to the best-fit vacuum system is $\mathcal{O}(100)$ cycles. As shown in the left panels of \cref{fig:discoverability}, lighter systems (including our benchmark system) than this will have an even larger dephasing (for example, exceeding 1000 cycles for $(m_1, m_2) = (10, 10^{-3}) \,\SI{}{\solarmass}$). We therefore conclude that over the parameter space of interest for light PBHs, the presence of a dark dress should be discoverable with observation times of 1 week or more.

We also show in \cref{fig:discoverability} the percentage SNR loss between the dephased system and the best-fitting vacuum system. For the system with $(m_1, m_2) = (10, 10^{-2}) \,\SI{}{\solarmass}$ described above, the SNR loss is roughly 40\%. This indicates that searching for dark dresses with vacuum templates will result in a large number of missed detections relative to a search based on dark dress templates. Given that the optimal matched-filtered SNR scales as $d_L^{-1}$ (see \cref{eq:SNR}), this SNR loss would correspond to a reduction of the detectable volume for these systems $\sim d_L^3$ by a factor of $(0.6)^3\sim 4.6$. The use of vacuum waveforms would substantially reduce the observable rate of large mass ratio PBH mergers. Looking again at lighter systems, the SNR loss increases further. For binaries with $(m_1, m_2) = (1, 10^{-3}) \,\SI{}{\solarmass}$, the SNR loss with CE approaches $80\%$, reducing the observable volume by a factor $\sim 100$, highlighting the importance of using dephased waveforms to effectively search for such systems.

\medskip

\paragraph{Measurability.} The one- and two-dimensional marginal posteriors for the intrinsic parameters of a dressed PBH system with benchmark masses of $m_1=\SI{1}{\solarmass}$ and $m_2=\SI{e-3}{\solarmass}$ observed with Einstein Telescope are shown in \cref{fig:post-et-week-month-year}. The true values of the parameters are shown by the dashed black lines, and the posteriors for week, month and year long signals are shown by the orange, green and red contours respectively. With even one week's worth of data, it will be possible to measure the slope of the density profile to precision $\gamma_s=2.22_{-0.05}^{+0.07}$,\footnote{
    We use error bars indicating the 68\% credible interval.
} and with a year's worth of data, $\gamma_s=2.25_{-0.01}^{0.01}$. The measured values for the other parameters are given in \Cref{tab:measurements}. 

We run the same parameter estimation for Cosmic Explorer, and find comparable errors on the parameters to Einstein Telescope in \Cref{fig:post-ce-week-month-year}. We note, however, that increasing the duration of the signal from one month to one year from merger does not improve the size of the contours for Cosmic Explorer because of the low-frequency cut-off of the noise curve - see \Cref{fig:trajectories-and-noise}.

In \Cref{fig:post-rho6-gammas} we highlight the one-dimensional marginal posteriors for $\rho_6$ and $\gamma_s$ to show the precision with which we can measure these parameters, as well as the improvement from increasing the duration of the signal. We stress, however, that one week's worth of data will be enough to measure these parameters with a few percent accuracy, and presents a less daunting data analysis challenge than year-long signals.

So far, we have included all of the intrinsic parameters of the system in the parameter estimation pipeline, specifically $\{ \gamma_s, \rho_6, \mathcal{M}, \log_{10} q \}$. However, for PBH systems, the value of $\gamma_s$ is fixed and the density normalization $\rho_6$ is uniquely determined by the central PBH mass, as described in \cref{sec:dephasing}. Assuming a PBH origin for the dephasing signal therefore allows us to reduce the parameter estimation problem to two dimensions -- $\{\rho_6, \mathcal{M} \}$ -- which substantially strengthens constraints, as we describe in \cref{app:2d_posteriors}.

\section{Discussion}
\label{sec:conclusions}

We have explored the prospects for detecting the presence of Dark Matter (DM) spikes around light primordial black hole (PBH) binaries with future ground-based gravitational wave (GW) observatories, such as Einstein Telescope (ET) and Cosmic Explorer (CE). We focus on a broad, well-motivated PBH mass function with $\fPBH \approx 0.066$, which is still allowed by current constraints from microlensing and the stochastic GW background. 

We can expect between 1 and 100 PBH mergers per $\mathrm{Gpc}^3$ volume per year, and an event rate for our specific benchmark system of $m_1=1 \pm 0.5\,{\mathrm M}_\odot$, $m_2=10^{-3}\,m_1$, of $\num{8e-3}$ and 0.3 events per year, observable with an SNR of at least 12 in each of ET and CE respectively. We note that if ET and CE are online simultaneously, the SNR threshold for each individual detector could be lowered, and hence the distance at which PBH mergers of a given mass are observable increases along with the event rate.

We find that for our benchmark system, the SNR lost if the waveform is matched with a vacuum template could be as large as $80\%$. This suggests that search strategies will need to take into account the effect of the DM spikes so as to avoid missing these signals. For larger $m_1$ than our benchmark system, the SNR loss will be smaller, so the system may be detectable with vacuum templates but would lead to biased parameter inference.

We show that using the correct model for parameter estimation, i.e.\ including the effects of the dark matter spike on the waveform, we can reconstruct the intrinsic parameters of the binary, the chirp mass and the mass ratio, as well as the parameters of the spike, the density normalisation and the power law of the dark matter density profile, to very good precision with one week's worth of data, as summarised in \Cref{tab:measurements}.

We find that Einstein Telescope can measure the dark matter spike parameters with better precision than Cosmic Explorer, owing to the lower frequency reach, which allows more cycles to be observed and hence more dephasing to accumulate. We note that there may be opportunities for multi-band observations of systems that overlap with the frequency ranges of both LISA and ET or CE~\cite{Cutler:2019krq}.

Since PBH binaries of these masses and mass ratios must be embedded in DM spikes (assuming they cannot make up all of the dark matter themselves), we conclude that in order to find these systems and measure their parameters correctly, the effect of dark matter on the phase of the inspiral must be taken into account in at least the parameter estimation process to avoid biased parameter inference. For some ranges of the parameter space (for example our benchmark system), the effect of the dark matter must be taken into account in order to detect the signal in the first place, as the SNR loss incurred by assuming the system is inspiralling in vacuum could be catastrophic.

We also emphasise that less extreme mass ratio mergers will also exist for our PBH formation scenario, which should be inspiralling in vacuum, since we expect their dark matter spikes to have been disrupted. Observations of these systems (potentially even with aLIGO/Virgo \cite{Nitz1,Nitz2}) would be a very strong indicator that the more extreme mass ratio systems are out there to be found, and would provide strong motivation for conducting a search for these exotic waveforms in the data.

These conclusions are drawn assuming that the search and inference will be conducted using matched filtering with template banks. However, considering the duration of the signals expected, techniques from continuous wave searches~\cite{lsc-cont} may be more suitable for searching for these signals that go through millions of cycles. Recently, Ref.~\cite{Guo:2022sdd} proposed the use of the Hough Transform to search for `mini-EMRIs', systems similar to those we consider here. The authors estimate that for a strain sensitivity similar to that of LIGO, a binary with masses $(10,\, 10^{-2})\,M_\odot$ may be detectable out to a distance of a few Mpc with this technique. This is roughly a factor of two less than the detectable distance we estimate in \cref{fig:snrs}, suggesting that the application of more realistic search strategies need not substantially degrade the detectability of the signals. 

In any case, it will be vital to understand the evolution of the frequency as a function of time for these systems and how it differs from the vacuum case. There is also potential for the use of machine learning to search the data for such long duration signals, in order to decrease the expense of computing the (many thousands of) waveforms of these systems directly.


While we have focused here on DM spikes around PBH binaries, many of the tools and conclusions apply also to other environmental effects. These include binaries embedded in accretion disks~\cite{PhysRevLett.124.251102} or `gravitational atoms' (clouds of light scalar fields bound to BHs)~\cite{Baumann:2018vus,Baumann:2021fkf,Baumann:2022pkl}. For these systems too, future ground-based observatories will provide exquisite sensitivity to slowly-accumulating dephasing effects. Our results suggest that $\mathcal{O}(\mathrm{weeks})$ of data would be required to extract useful physical information from such systems, though a detailed study of this -- and of whether the environmental effects from these different sources can be distinguished -- we leave for future work. 

\section*{Acknowledgements}

We thank Thomas Edwards, Andrew Gow, Samaya Nissanke and Ville Vaskonen for useful conversations.

P.C. acknowledges funding from the Institute of Physics, University of Amsterdam.
A.C. received funding from the Netherlands eScience Center (grant number ETEC.2019.018) and the Schmidt Futures Foundation.
B.J.K.\ thanks the Spanish Agencia Estatal de Investigaci\'on (AEI, Ministerio de Ciencia, Innovación y Universidades) for the support to the Unidad de Excelencia Mar\'ia de Maeztu Instituto de F\'isica de Cantabria, ref. MDM-2017-0765.
We acknowledge Santander Supercomputing support group at the University of Cantabria who provided access to the supercomputer Altamira at the Institute of Physics of Cantabria (IFCA-CSIC), member of the Spanish Supercomputing Network, for performing simulations/analyses.

\appendix

\section{GW spectrum from coalescing BHs}
\label{app:GWspectrum}

For the interested reader, we here give explicitly expressions for the GW spectrum of coalescing black holes, which appears in the calculation of the stochastic GW background in \cref{sec:PBHconstraints}.

The energy emitted by coalescing BHs with masses $m_1$ and $m_2$ in the GW frequency range $[f, f + \diff f]$ is given by~\cite{Raidal_2017,Chernoff:1993th,Zhu:2011bd}:
\begin{gather*}
\frac{\diff E_{GW}}{\diff f} = \mathcal{C}
\begin{cases}
  f^{-\frac{1}{3}} & \text{for }f<f_1\,,\\    
  \frac{f^\frac{2}{3}}{f_1} & \text{for }f_1 \le f<f_2\,,\\  
    \frac{ff_3^4}{(f_1f_2^\frac{4}{3}(4(f - f_2)^2 + f_4)^2)^2} & \text{for }f_2 \le f<f_3\,,\\  
        0 & \text{for }f_3 \le f<f_4\,,
\end{cases}
\end{gather*}
where the prefactor is $\mathcal{C} = (\pi G)^\frac{2}{3}M^\frac{5}{3}\eta $.
The frequency window limits are given numerically by~\cite{Ajith:2007kx}:
\begin{equation}
    f_j=\frac{(a_j\eta^2+b_j\eta+c_j)c^3}{\pi G(m_1+m_2)}
\end{equation}
where e.g. $a_j$ denotes the $j$-th index of the following arrays:
\begin{align}
a&=(0.2971,0.59411,0.84845,0.50801),\\
b&=(0.04481,0.089794,0.12848,0.077515),\\
c&=(0.095560,0.19111,0.27299,0.022369),
\end{align}
and $\eta=m_1m_2/(m_1+m_2)^2$.

\section{HaloFeedback Validation}
\label{sec:HalofeedbackValidation}

Here, we present some validation tests which were performed for the $f(t)$ parametrization presented in Ref.~\cite{Coogan:2021uqv} and used in Sec.~\ref{sec:measurability} of this work. 

\begin{figure}[tb]
\begin{center}
    \includegraphics[width=0.49\textwidth]{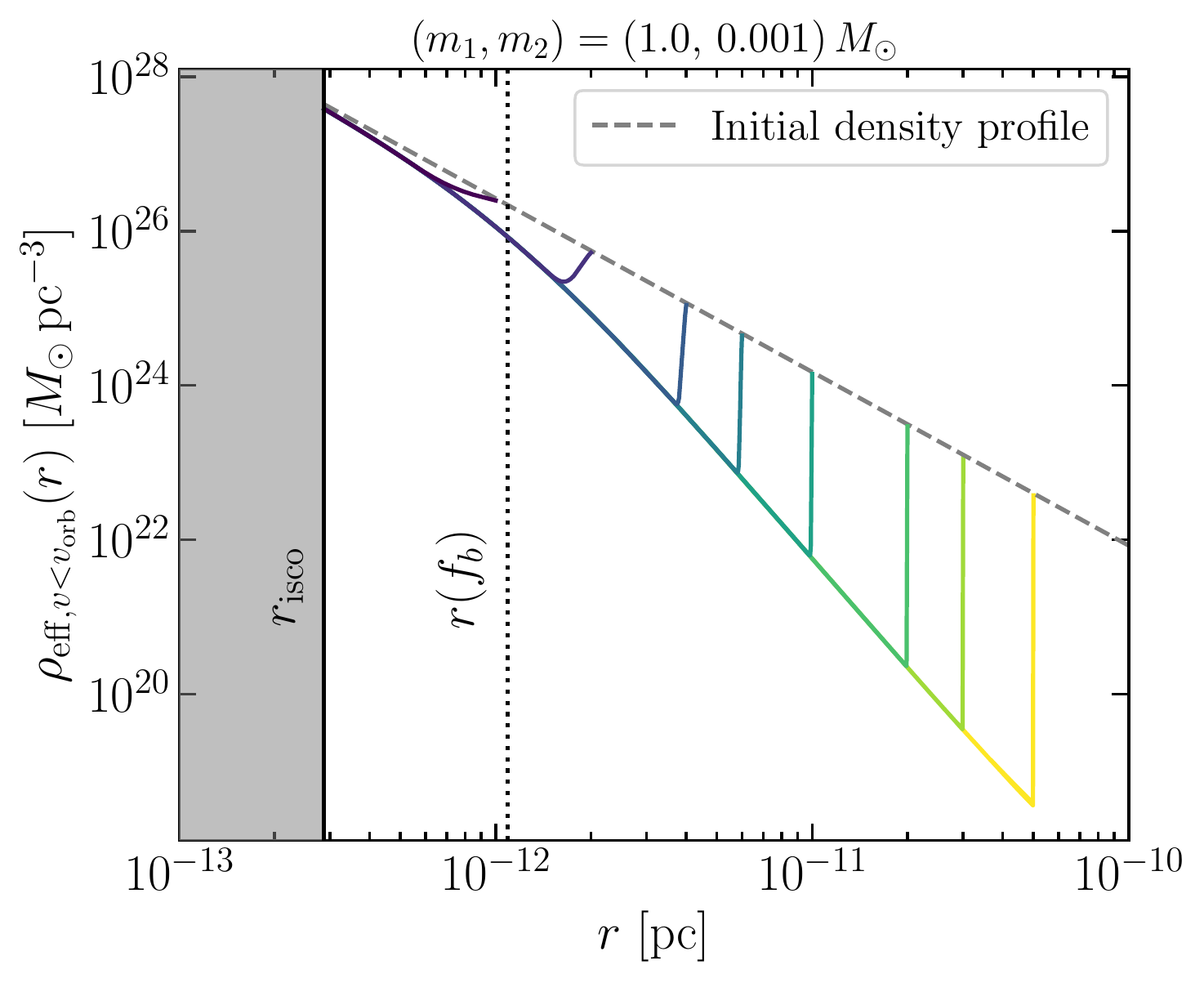}
        \caption{\textbf{Effective density profile of the DM spike, obtained using \texttt{HaloFeedback}.} Each colored line corresponds to a simulation beginning at a different initial PBH binary separation, while the diagonal dashed line shows the unperturbed DM density profile. The behaviour of the system at small radii is independent of the initial separation.}
        \label{fig:HaloFeedback_densities}
    \end{center}
\end{figure}

We begin by checking that the formalism developed in Ref.~\cite{Kavanagh:2020cfn} behaves well for the low BH masses we consider here. 
In \Cref{fig:HaloFeedback_densities}, we show the effective density profile extracted from simultaneously evolving the separation of the PBH binary and the DM spike distribution function (as summarized in \Cref{sec:dephasing} and implemented in the \texttt{HaloFeedback} code~\cite{HaloFeedback}). The effective density is the instantaneous DM density experienced by the smaller inspiraling BH when it reaches an orbital separation $r$. For reference, the orbital reference corresponding to the break frequency $f_b$ for this system is shown as a vertical dotted line.

The qualitative behaviour of the effective density matches that observed in the heavier systems presented in Ref.~\cite{Coogan:2021uqv}. Because of energy injected by the BH into the spike, the spike is rapidly depleted and thus the effective density is smaller than the initial, unperturbed density (grey dashed line). As the GW inspiral continues (from large $r$ to small $r$), the timescale for depletion of the spike eventually becomes longer than the timescale for the GW inspiral, and the effective density converges to the initial unperturbed profile. Lines of different colors in \Cref{fig:HaloFeedback_densities} correspond to simulations which were started at different initial BH separations. After an initial depletion phase, each of these density profiles converges to the same behaviour, indicating that we do not need to explicitly specify the initial separation of the binary at formation (as long as this is larger than the radius at which the binary enters the GW observing band). 

\begin{figure}[tb]
\begin{center}
    \includegraphics[width=0.48\textwidth]{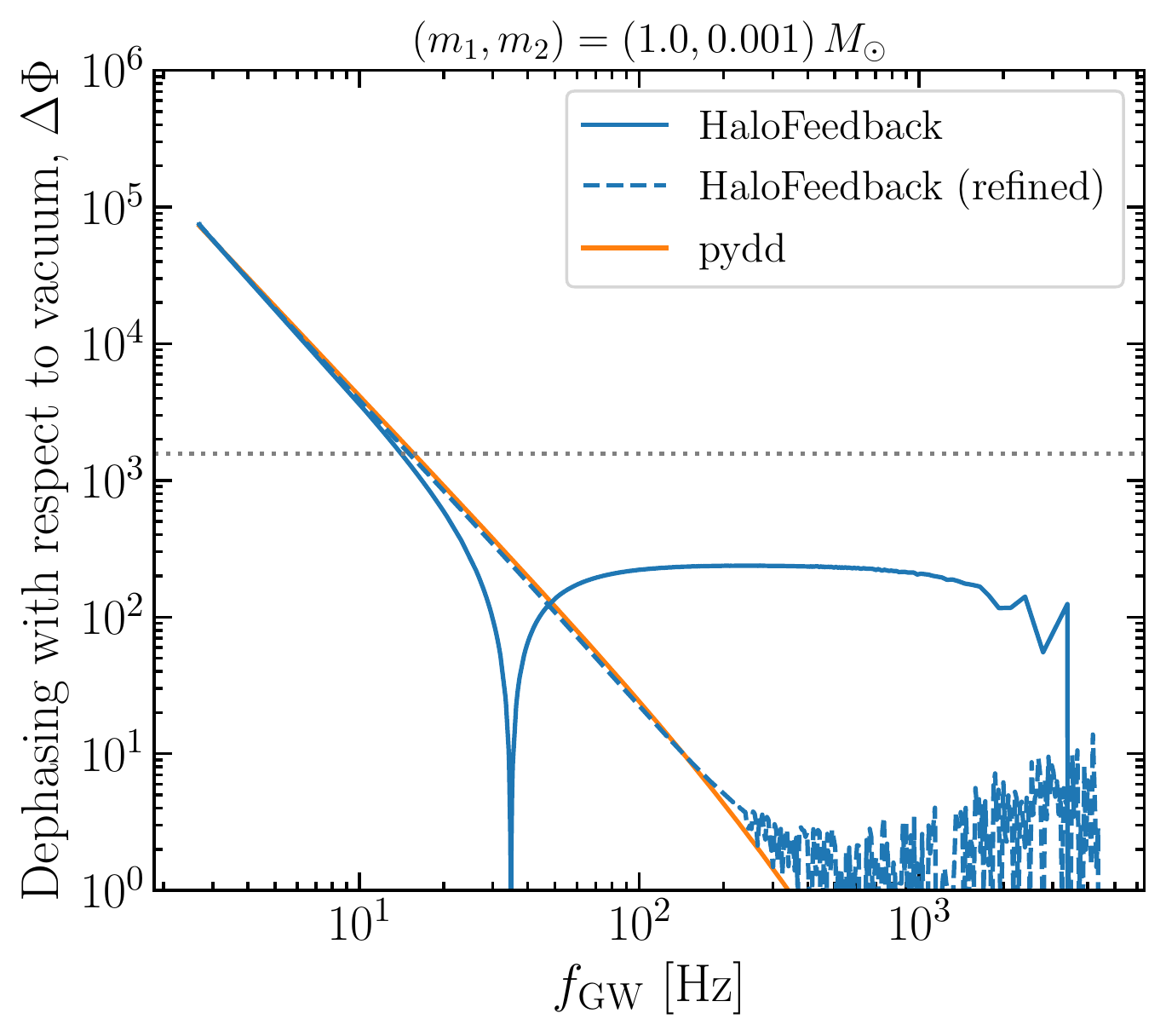}
        \caption{\textbf{DM-induced dephasing with respect to the vacuum for different modeling approaches.} The solid blue line shows the dephasing modeled using the \texttt{HaloFeedback} code~\cite{HaloFeedback}, solving simultaneously the inspiral and feedback on the DM spike. For computational ease, the maximum step size is set to 100 orbits (corresponding to a phase given by the horizontal dotted line). The dashed blue line is obtained assuming a static DM spike with effective density profile extracted from the full \texttt{HaloFeedback} run, allowing for a shorter timestep. This matches closely the output of \texttt{pydd}~\cite{pydd} (solid orange line), which implements the analytic fit to the phase $\Phi(f)$ used in Sec.~\ref{sec:measurability} of this work and first presented in Ref.~\cite{Coogan:2021uqv}.}
        \label{fig:HaloFeedback_error}
    \end{center}
\end{figure}

Having verified that the behaviour of \texttt{HaloFeedback} is sensible for these light systems, we can now compare these simulations against the phase parametrization $\Phi(f)$ which we use for parameter estimation in this work. From the trajectories obtained in these simulations, we can calculate the evolution of the GW phase with frequency and compare this with the corresponding predictions from the analytic phase parametrization. In \Cref{fig:HaloFeedback_error}, we plot the dephasing with respect to the vacuum case for PBH binaries with masses $(m_1, \,m_2) = (1.0,\, 10^{-3}) \,M_\odot$. The full \texttt{HaloFeedback} simulations are shown in solid blue, using a maximum timestep of 100 orbits (corresponding to a phase given by the horizonal dotted line). We use this simulation to extract the effective DM density profile and resimulate assuming a \textit{static} spike with this density profile. This allows us to substantially reduce the maximum timestep and therefore resolve the dephasing much closer to the merger (dashed blue line). These results are matched closely by the output of the analytical parametrization, as implemented in \texttt{pydd}~\cite{pydd}, down to the level of a few cycles. 

Despite being calibrated on BH masses a few orders of magnitude larger than those being considered here, the analytic parametrization maintains percent-level accuracy over many years of the inspiral. We have also verified this behaviour for various mass ratios and total masses of the light PBH systems.

\section{Posteriors for fixed \texorpdfstring{$\gamma_s$}{gamma\_s}}
\label{app:2d_posteriors}

\begin{figure}
    \centering
    \includegraphics[width=0.48\textwidth]{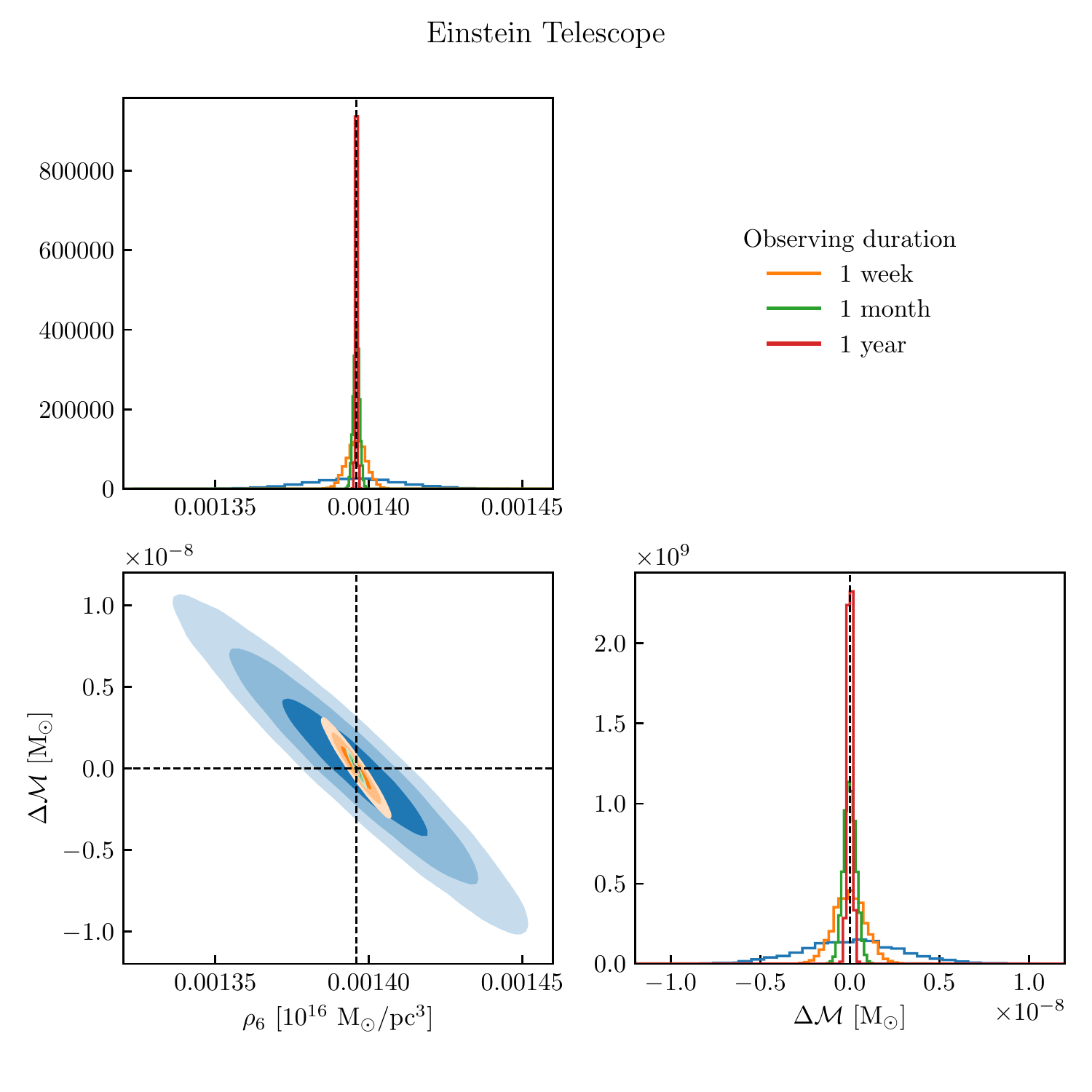}
    \caption{\textbf{One- and two-dimensional marginal posteriors for a subset of the intrinsic parameters of PBH dark dresses, keeping $\gamma_s$ and $q$ fixed, observed with Einstein Telescope for times ranging from one week to one year.} The dashed black lines indicate the true parameter values. The black hole masses are set to the benchmark values of \SI{1}{\solarmass} and \SI{e-3}{\solarmass}, and $\Delta \mathcal{M}$ is the difference between the measured and true chirp mass ($\sim \SI{1.585e-2}{\solarmass}$). The distances for the systems are set so each is detected with an \gls*{snr} of 12. The contours indicate the \SI{65}{\percent}, \SI{95}{\percent} and \SI{99.7}{\percent} credible regions.}
    \label{fig:fixed-gamma-et}
\end{figure}

\begin{figure}
    \centering
    \includegraphics[width=0.48\textwidth]{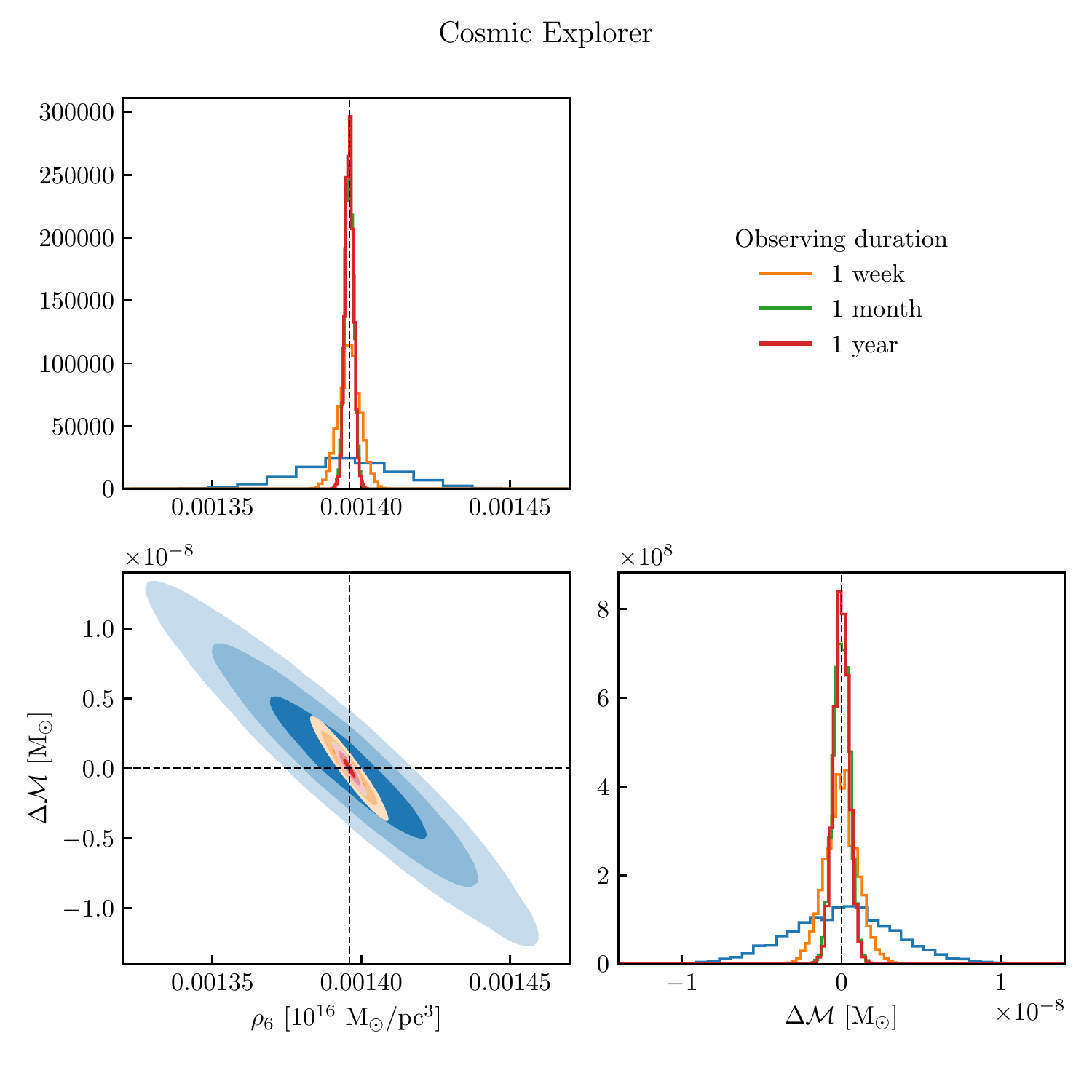}
    \caption{\textbf{Marginal posteriors for a PBH dark dress with $\gamma_s$ and $q$ fixed as in \Cref{fig:fixed-gamma-et}, but instead observed with Cosmic Explorer.}}
    \label{fig:fixed-gamma-ce}
\end{figure}

In the main body of the paper, we have performed parameter estimation for all four intrinsic parameters of the dark dress systems: $\gamma_s$, $\rho_6$, $\mathcal{M}$ and $q$. However, it is also interesting in the case of PBHs to make use of the fact that there is a concrete prediction for the slope of the spike $\gamma_s=9/4$, as well as that $\rho_6$ is directly related to the mass of the primary black hole $m_1$. With this in mind, we fix $\gamma_s=9/4$ and perform inference over the parameters $\rho_6$ and $\mathcal{M}$, using \Cref{eq:pbhspike} and \Cref{eq:pbhspike_rho6} to fix the mass ratio $q$. This is a strong assumption on two fronts: firstly it  assumes that the system is definitely primordial (reasonable if the primary mass is below $\SI{1.4}{\solarmass}$) and that the predictions from analytic calculations and simulations are correct and have no scatter, and secondly it assumes that the spike has not been at all disrupted down to redshifts less than one, such that the density profile slope and relationship between the normalisation and $m_1$ remains intact. While we do not foresee this type of parameter estimation being implemented for the discovery of the first system of this type, it demonstrates how well we can predict the parameters of the system if we are confident that we are observing one of a (perhaps previously confirmed) population of primordial black holes with pre-existing evidence that the density profile slope of these systems is $\gamma_s=9/4$.

The resulting posteriors are shown in \Cref{fig:fixed-gamma-et} for an analysis with Einstein Telescope and in \Cref{fig:fixed-gamma-ce} for an analysis with Cosmic Explorer. Unsurprisingly, in both cases the posteriors are extremely narrow even with just one day of observations before coalescence. The chirp mass can be measured to within approximately $3\times10^{-9}$ of the true value, and the density normalisation $\rho_6$ to better than \SI{1}{\percent} precision with \SI{99.7}{\percent} confidence.

\bibliography{main.bib}

\end{document}